\documentclass[10pt,twocolumn,letterpaper]{article}

\usepackage[pagenumbers]{cvpr} % To force page numbers, e.g. for an arXiv version

\input{preamble}

\definecolor{cvprblue}{rgb}{0.21,0.49,0.74}
\usepackage[pagebackref,breaklinks,colorlinks,citecolor=cvprblue]{hyperref}

\def\paperID{11218} % *** Enter the Paper ID here
\def\confName{CVPR}
\def\confYear{2024}

\title{MDSC: Towards Evaluating the Style Consistency Between Music and Dance}

\author{Zixiang Zhou$^\dagger$\\
Xiaobing.AI\\
{\tt\small zhouzixiang@xiaobing.ai}
\and
Weiyuan Li$^\dagger$\\
Xiaobing.AI\\
{\tt\small liweiyuan@xiaobing.ai}
\and
Baoyuan Wang\\
Xiaobing.AI\\
{\tt\small wangbaoyuan@xiaobing.ai}
}

\begin{document}
\twocolumn[{
\maketitle
\begin{center}
    \includegraphics[width=0.98\textwidth]{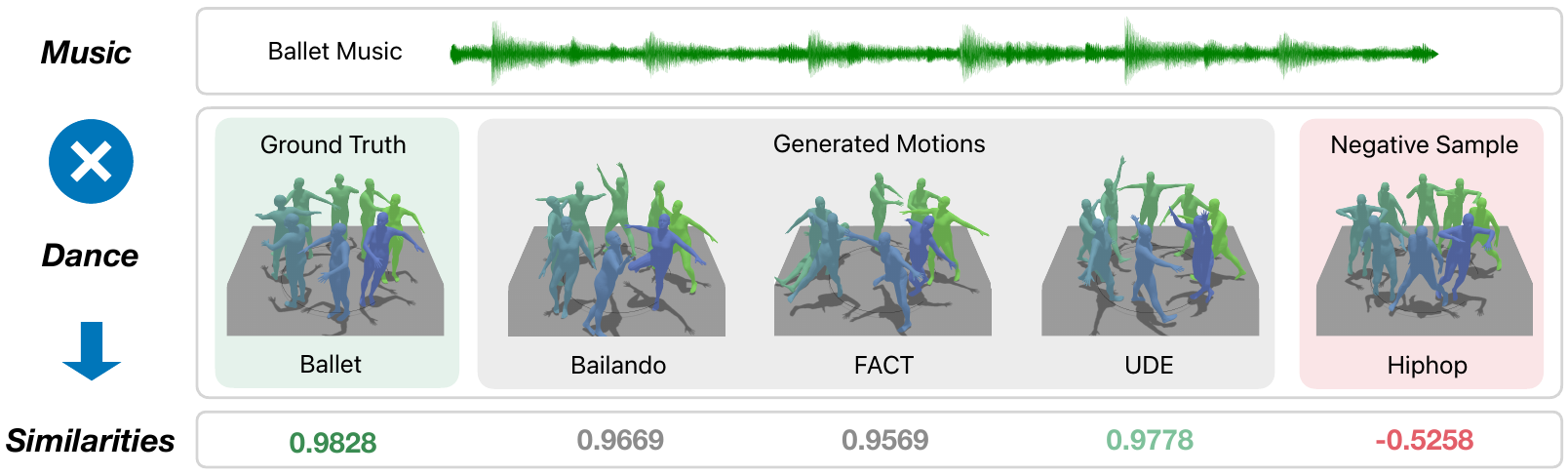}
    \captionof{figure}{Given a music stream and five dance motion sequences, our model is able to measure style consistency between dance motion and music.}
    \label{fig:teaser}
\end{center}
}]

\def\thefootnote{$\dagger$}\footnotetext{These authors contributed equally to this work}

% \maketitle
\begin{abstract}
   We propose \textbf{MDSC}(Music-Dance-Style Consistency), the first evaluation metric that assesses to what degree the dance moves and music match. Existing metrics can only evaluate the motion fidelity and diversity and the degree of rhythmic matching between music and dance. \textbf{MDSC} measures how stylistically correlated the generated dance motion sequences and the conditioning music sequences are. We found that directly measuring the embedding distance between motion and music is not an optimal solution. We instead tackle this through modeling it as a clustering problem. Specifically, 1) we pre-train a music encoder and a motion encoder, then 2) we learn to map and align the motion and music embedding in joint space by jointly minimizing the intra-cluster distance and maximizing the inter-cluster distance, and 3) for evaluation purposes, we encode the dance moves into embedding and measure the intra-cluster and inter-cluster distances, as well as the ratio between them. We evaluate our metric on the results of several music-conditioned motion generation methods \cite{li2021ai}\cite{siyao2022bailando}\cite{zhou2023ude}, combined with user study, we found that our proposed metric is a robust evaluation metric in measuring the music-dance style correlation. 
   % The code is available at: \href{https://github.com/zixiangzhou916/MDSC}{https://github.com/zixiangzhou916/MDSC}
\end{abstract}    
\section{Introduction}
\label{sec:intro}

\setlength{\parindent}{1em} Synthesizing realistic human motion sequences has made remarkable progress in recent years. It is now possible to synthesis human motion sequence from natural language descriptions \cite{petrovich2021action}\cite{Guo_2022_CVPR}\cite{tevet2022human}\cite{zhou2023ude}, or from music \cite{siyao2022bailando}\cite{zhou2023ude}\cite{tseng2023edge}. Despite these achievements in motion synthesis, less progress has been made in terms of proper evaluation metrics. Further improvements on motion generation could hardly be made without comprehensive and fine-grained evaluation metrics. Therefore, it is vital for the community to develop proper metrics on evaluating the outcomes of the human motion synthesis.\par

\setlength{\parindent}{1em} Since conditioned human motion generation is a typical one-to-many mapping problem, there are multiple aspects critical for the evaluation metrics to take into consideration. \cite{ye2022human} summarized there are four major categories to be considered when evaluating the motion generation, and they are: 1) \textbf{fidelity}: it measures the quality and smoothness of the generated motion sequences, 2) \textbf{diversity}: it measures how diverse the synthesized motions are given same driving source, 3) \textbf{condition consistency}: it measures how correlated the generated motion sequences and the driving sources are in terms of semantic meaning, rhythmic pattern, or style, and 4) \textbf{user study}: it measures the motion generation results from human's perspective, which is more subjective compared with other three categories.\par

\setlength{\parindent}{1em} Various metrics have been proposed to evaluate the synthesized motion sequence on text conditioned scenario\cite{ahuja2019language2pose}\cite{ghosh2021synthesis}\cite{habibie2022motion}\cite{guo2020action2motion}\cite{lu2022action}\cite{lee2023multiact}\cite{tevet2022human}\cite{chen2023executing}\cite{zhang2023t2m}\cite{zhou2023ude}, and these methods cover the four major categories of evaluation. Specifically, 1) \cite{ahuja2019language2pose}\cite{ghosh2021synthesis}\cite{lu2022action} focus on evaluating the motion fidelity, 2) and \cite{huang2020dance}\cite{habibie2022motion}\cite{guo2022tm2t}\cite{tevet2022human}\cite{sun2022you} propose to evaluate the motion diversity, 3) \cite{guo2020action2motion}\cite{lu2022action}\cite{lee2023multiact}\cite{tevet2022human}\cite{chen2023executing}\cite{zhang2023t2m}\cite{zhou2023ude} propose metrics on measuring how semantically consistent the generated motion sequences and the conditioning text descriptions are, and 4) \cite{petrovich2022temos}\cite{tevet2022motionclip}\cite{zhou2023ude} propose protocols on evaluating the quality of motion synthesis from subjective perspective.\par 

\setlength{\parindent}{1em}While the metrics for assessing motion fidelity and diversity are condition-independent, the metrics for condition consistency are condition-specific. The definition of consistency for music-conditioned is quite different from text-conditioned. The music-motion consistency are two folds, on one hand, rhythmic consistency plays a vital role in evaluating the music driven motion quality\cite{lee2019dancing}\cite{huang2020dance}\cite{alexanderson2022listen}\cite{tseng2023edge}\cite{au2022choreograph}\cite{li2021ai}\cite{siyao2022bailando}, on the other hand, whether the dance motion style is consistent with the music style is also critical. Unlike the definition of text-conditioned consistency, music-conditioned consistency is much more relaxed. Specifically, text-to-motion is a one-to-many mapping. For example, the text description '\textit{a person is running.}' could be mapped to various motion sequence, as long as they demonstrate the same semantic meaning, but these motion sequences cannot be mapped to the description '\textit{a person is walking.}'. 
For music-to-motion, however, it is a many-to-many mapping, which means a ballet style music could be mapped to various ballet style dances, and reversely, these \textit{ballet style} dances could also be mapped to various \textit{ballet style} music. These dances could be choreographically different, and the music arrangement styles could also vary a lot.\par

\setlength{\parindent}{1em}To measure the consistency between music and dance, we don't measure the embedding similarity between music clip and motion sequence, as \cite{messina2023text} does for text-to-motion scenario. Instead, we model it as an embedding clustering problem. We use two encoders to obtain embedding from music and motion sequences, respectively, and cluster the embedding from stylistically consistent music-motion pair into same cluster. Meanwhile, we push the clustering centers apart from each other to maximize the inter-cluster embedding distance. When evaluation, 1) we can only encode dance moves and measure the intra-cluster and inter-cluster distances between encoded embedding and clustering centers embedding, or 2) we can encode both music and dance and measure the distance between their embedding.\par

\setlength{\parindent}{1em}Our contributions are three folds: 1) We define the music-to-motion style consistency and model it as a quantifiable problem. 2) We propose the first, to the best of our knowledge, music-to-motion style consistency evaluation model as a metric, and conduct comprehensive experimental analysis to validate the effectiveness of our method. And 3) we provide baselines as measurements of music-to-motion consistency for future research.\par
\section{Related Work}
\label{sec:relatedwork}

\paragraph{Music Representation Learning}Music representation learning has been widely studied in music auto-tagging and classification\cite{choi2016automatic}\cite{choi2019zero}, music retrieval\cite{won2021multimodal}\cite{doh2023toward} and music understanding\cite{mccallum2022supervised}.\par 

\setlength{\parindent}{1em}For music auto-tagging and classification, the task is to obtain various attributes from music streams, including the music genres, the rhythmic traits, the musical moods, etc. Typically, a music stream is likely to be categorized into multiple classes, making it difficult to define the categories of music attributes and also difficult in classifying the music streams\cite{won2021music}. There are two major categories of learning paradigms, namely, supervised and self-supervised learning in auto-tagging and classification. 1) For supervised learning paradigm, various types of neural networks architectures have been studied and proven effective in learning features from labeled datasets\cite{pons2019musicnn}\cite{won2020evaluation}\cite{won2021semi}. However, labeled datasets are costly to obtain, and the categories are unlimited, making it an open vocabulary problem. 2) As an alternative, self-supervised learning paradigm has numerous advantages against supervised counterpart. Instead of learning from labeled data, self-supervised paradigm attempts to learn patterns from tremendous unlabeled data. Contrastive learning\cite{oord2018representation} is an effective learning technique, and multiple studies have been proposed and proven effective in effective musical representations from unlabeled data\cite{chen2020simple}\cite{spijkervet2021contrastive}\cite{mccallum2022supervised}.\par

\setlength{\parindent}{1em}Music retrieval and understanding is normally involved with multimodality representation learning. For example, text-based music retrieval\cite{won2021emotion}\cite{chen2022learning}\cite{huang2022mulan}\cite{doh2023toward} attempts to learn a joint representation space between music streams and natural language descriptions. For a semantically aligned pair of music and description, their embedding are pulled together in the joint space, while for misaligned pairs, their embedding are pushed apart. Similarly, image-based or video-based music retrieval and understanding is designed to learn the joint representation space between acoustic and visual modalities. For example, \cite{yariv2023audiotoken}\cite{girdhar2023imagebind} attempts to correlate visual content and acoustic content using contrastive learning paradigm. The learnt representation could be effectively employed for image-based music retrieval and, in reverse, audio-based image retrieval.\par
\vspace{-3mm}

\paragraph{Motion Representation Learning} Motion representation learning could be categorized into motion recognition \cite{liu2020disentangling}\cite{chi2022infogcn}\cite{zhou2023learning}, and motion understanding \cite{petrovich2023tmr}\cite{kulal2021hierarchical}\cite{endo2023motion}.\par

\setlength{\parindent}{1em}Motion recognition, or action recognition, is the task that estimates the category of the query motion sequence. Typically, given a query of motion sequence, which is normally represented by 3D skeleton joints\cite{chi2022infogcn} or rotation parametric prior\cite{tevet2022motionclip}, one or multiple action categories are estimated to describe the motion. Mostly, these are trained on predefined set of action categories using supervised paradigm \cite{liu2020disentangling}\cite{xu2022topology}\cite{chi2022infogcn}\cite{zhou2022hypergraph}\cite{zhou2023learning}. However, predefined action categories are limited and normally able to describe short and simple actions, open vocabulary action recognition is an alternative solution to this\cite{tevet2022motionclip}\cite{petrovich2023tmr}\cite{messina2023text}. Instead of learning the direct mapping between actions and labels, these methods attempts to learn a joint representation space that align the actions and descriptions, and retrieval based strategy is employed for open vocabulary recognition.\par

\begin{figure*}[h]
    \centering
    \includegraphics[width=\textwidth]{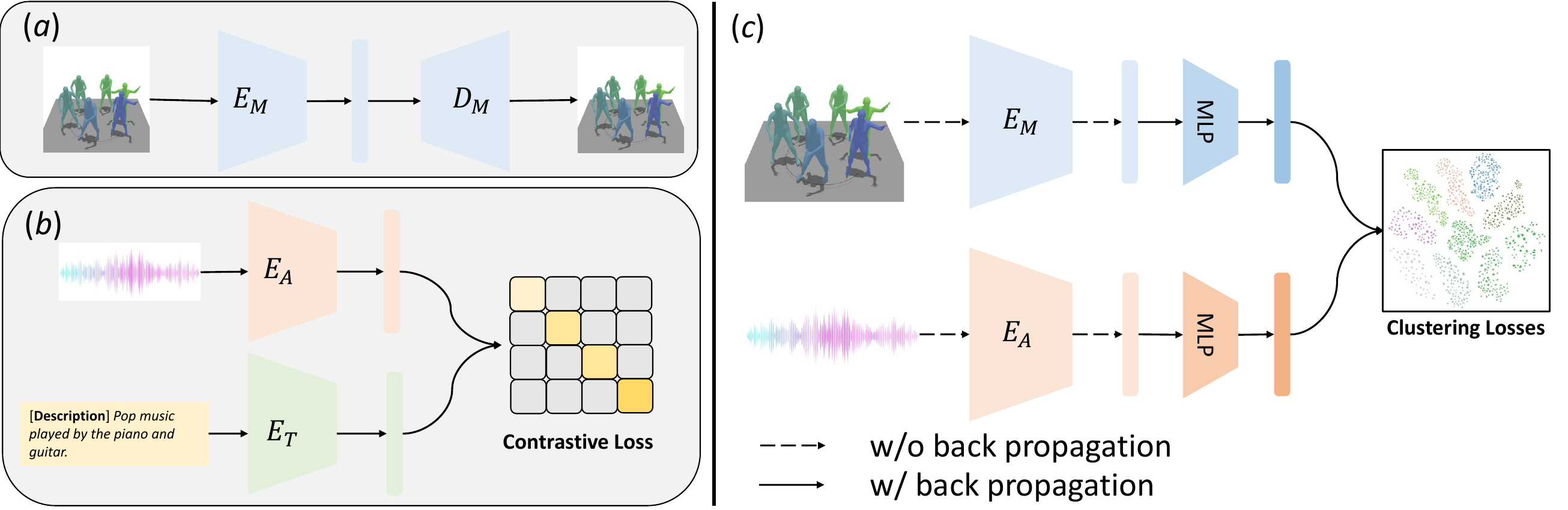}
    \caption{\textbf{Pipeline of Music-Dance Style Consistency} (a) We train a motion auto-encoder supervised by reconstruction loss, and use the encoder as $\textit{E}_M$. (b) We use the pretrained music encoder in \cite{messina2023text} as our $\textit{E}_A$. (c) Given batch of motion sequence and music streams as input, our method uses pretrained motion encoder $\textit{E}_M$ and music encoder $\textit{E}_A$ to obtain their embedding. Instead of pulling paired motion embedding and audio embedding closer and push unpaired apart, we attempt to cluster style-consistent motion embedding and music embedding into same cluster, while inconsistent embedding are clustered into different clusters. At this stage, only the light-weight \textbf{MLPs} are trainable. The \textbf{dotted arrow} means no back-propagation is applied, while \textbf{solid arrow} means back-propagation is applied.}
    \label{fig:overview}
\end{figure*}

\setlength{\parindent}{1em}Motion understanding is a more open problem compared with recognition task, and typically applies to complex and long action sequence. In addition to estimating the action categories, these tasks also attempts to reason from the action. For example, in \cite{kulal2021hierarchical}\cite{endo2023motion}, they attempt to understand the action sequence from global level to local level. This does not only require alignment between action representation and description representation, but also alignment between pose or body part representation and word or phrase representation.\par
\vspace{-3mm}

\paragraph{Music-Motion Consistency} Current music-motion consistency studies focus on assessing the rhythmic consistency between music and motions. These studies \cite{lee2019dancing}\cite{huang2020dance}\cite{sun2022you}\cite{li2022danceformer}\cite{au2022choreograph}\cite{siyao2022bailando} asses the rhythmic consistency by beat alignment score, which measures the degree of motion kinetic beats and the musical beats are aligned. Although motion kinetic beats are defined in different aspects, they assume that high music-motion consistency means better alignment between kinetic beats and musical beats, regardless the music styles and dance choreography. However, the evaluation of music-motion consistency is a non-trivial problem and could hardly be defined from single aspect. Existing studies are insufficient to evaluate the correlation between music and dance motion objectively and comprehensively.\par
\vspace{-3mm}

\paragraph{Summary}We found that although numerous research efforts have been made in understanding music and human motion, respectively, few work is proposed to bring them together. Although few work measure the consistency in terms of rhythmic matching, these methods are insufficient in evaluating the style consistency between music and dance moves. Hence, we propose a novel approach to align music and motion semantically, and show that it could be used as an evaluation metric for music-conditioned motion generation task.\par

\section{Method}

We show the overview pipeline of our method in Fig. \ref{fig:overview}, which contains a pretrained motion encoder $\textit{E}_M$, a pretrained music encoder $\textit{E}_A$, and two light-weight MLPs. The detail of each module is described in following sections.\par

\subsection{Music Encoder} We use pretrained music encoder in \cite{messina2023text}, which is a modified version of music tagging transformer\cite{won2021semi}. The pretraining scheme is shown in Fig. \ref{fig:overview}(b), where a pretrained text encoder and a modified audio encoder are adopted to obtain music and text representation, respectively. Following typical CLIP training paradigm\cite{radford2021learning}, the encoders are trained to maximize the similarity of embedding of music and text aligned pairs, and to minimize the similarity of misaligned pairs. Readers are refered to \cite{messina2023text} for details.\par

\subsection{Motion Encoder} As shown in Fig. \ref{fig:overview}(a), 
we train a motion auto-encoder and adopt the encoder part as a good motion representation encoding prior. 
Given a motion sequence denoted as: $x\in\mathbbm{R}^{{T}^\times{c}}$, 
where $T$ and $c$ are temporal length and dimension per frame. 
We use an encoder $\mathcal{E}_{M}(\cdot)$ to obtain embedding $z_{M}$ from input motion sequence as: $z_{M} = \mathcal{E}_{M}(x)$, and a decoder $\mathcal{D}_{M}(\cdot)$ to reconstruct motion sequence $\tilde{x}$ from $z$ as: $\tilde{x} = \mathcal{D}_{M}(z_M)$. 
Therefore, the motion auto-encoder process is modeled as Eq. \ref{eq:auto-encoder}:
\vspace{-3mm}

\begin{equation}
    \tilde{x} = \mathcal{D}_M(\mathcal{E}_M(x))
    \label{eq:auto-encoder}
\end{equation}
\vspace{-3mm}

And the motion auto-encoder is training by minimizing the reconstruction error depicted as Eq. \ref{eq:auto-encoder-loss-fn}:
\vspace{-3mm}

\begin{equation}
    \mathcal{L}_{rc} = \|\tilde{x} - x\|
    \label{eq:auto-encoder-loss-fn}
\end{equation}
\vspace{-3mm}

After the auto-encoder is trained, we adopt its encoder part as our motion encoder.\par

\subsection{Music-Dance Style Alignment}Given a pair of dance motion sequence $x_{M}$ and music clip $x_{A}$, as shown in Fig. \ref{fig:overview}(c), we first obtain their embedding $z_{M}\in\mathbbm{R}^{1\times{c_M}}$ and $z_{A}\in\mathbbm{R}^{1\times{c_A}}$ using the pretrained audio encoder and motion encoder, respectively, where $c_{M}$ and $c_{A}$ are the dimension of motion embedding and music embedding. This is denoted as $z_{M}=\mathcal{E}_{M}(x_{M})$ for motion encoding and $z_{A}=\mathcal{E}_{A}(x_{A})$ for music encoding.\par

\begin{figure}[h]
    \centering
    \includegraphics[width=\linewidth]{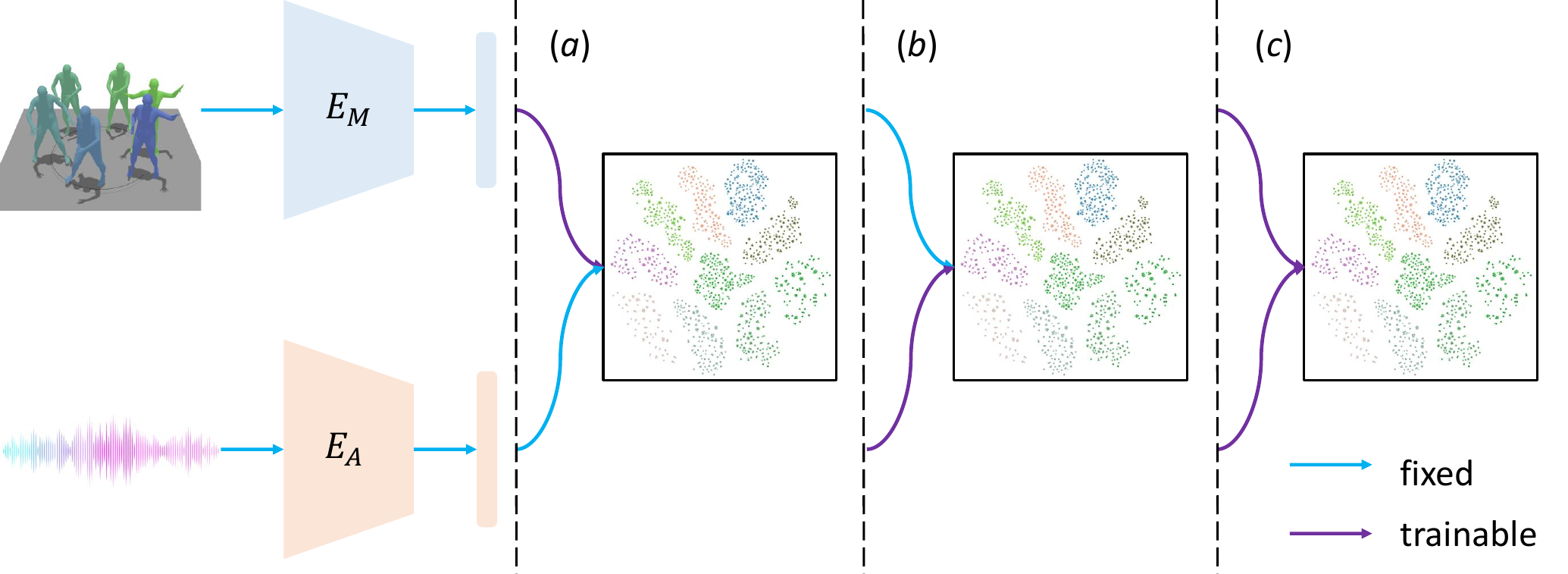}
    \caption{\textbf{Variants of Design in Aligning Cross-Modality Embedding}Alignment of motion embedding and music embedding in three different approaches. (a) Fix music embedding and align motion embedding. (b) Fix motion embedding and align music embedding. (c) Align both music and motion embedding to joint space. $\color{cyan}{\rm Fixed}$, $\color{violet}{\rm Trainable}$.}
    \label{fig:variants}
\end{figure}

\setlength{\parindent}{1em}Because the audio encoder $\mathcal{E}_{A}(\cdot)$ and motion encoder $\mathcal{E}_{M}(\cdot)$ are adopted from pretrained models, and these pretrained models are trained using different dataset and under different settings, their output embedidng are not necessarily aligned in latent space. To align the embedding from different encoders in latent space, there are three possible design options shown in Fig. \ref{fig:variants}, and we will discuss them in detail as follow:\par

\vspace{-3mm}
\paragraph{Align Motion Embedding to Music Embedding}We assume the audio encoder is powerful enough to extract stylistically meaningful feature embedding from music sequence. Therefore, as long as we we can align paired motion embedding to music embedding, the music-to-dance style consistency could be measured. This is shown in Fig. \ref{fig:variants}(a), where both motion encoder $\mathcal{E}_{M}(\cdot)$ and audio encoder $\mathcal{E}_{A}(\cdot)$ are fixed. To align motion embedding $z_{M}$ to music embedding $z_{A}$, we adopt a MLP to project the motion embedding to audio embedding as Eq. \ref{eq:align-m2a}.\par
\vspace{-3mm}

\begin{equation}
    \textit{f}_{{M}\rightarrow{A}}(z_{M}): \mathbbm{R}^{1\times{c_M}} \rightarrow \mathbbm{R}^{1\times{c_A}}
    \label{eq:align-m2a}
\end{equation}
\vspace{-10mm}

\paragraph{Align Music Embedding to Motion Embedding}Similarly, we assume the motion encoder obtains dance motion embedding with rich style information. In this case, aligning the music embedding to motion embedding will suffice, presumably, in measuring the dance-music style consistency. Again, for this case, both motion and music encoders are kept fixed during training, while a MLP is injected to project music embedding to motion embedding as Eq. \ref{eq:align-a2m}. Visual illustration is given in Fig. \ref{fig:variants}(b).\par
\vspace{-3mm}

\begin{equation}
    \textit{f}_{{A}\rightarrow{M}}(z_{A}): \mathbbm{R}^{1\times{c_A}} \rightarrow \mathbbm{R}^{1\times{c_M}}
    \label{eq:align-a2m}
\end{equation}
\vspace{-10mm}

\paragraph{Align Music and Motion Embedding in Joint Space}For this design option, we assume neither pretrained motion encoder nor audio encoder obtain stylistically representative embedding for style consistency evaluation. Therefore, we attempt to learn a joint embedding space which is representative for the style consistency measurement. In this case, as shown in Fig. \ref{fig:variants}(c), we adopt two MLPs, one for motion embedding and another for audio embedding respectively, to project the motion and audio embedding to joint space. We denote this process as Eq. \ref{eq:align-a2j} for audio embedding, and Eq. \ref{eq:align-m2j} for motion embedding.\par
\vspace{-3mm}

\begin{equation}
    \textit{f}_{{A}\rightarrow{J}}(z_{A}): \mathbbm{R}^{1\times{c_A}} \rightarrow \mathbbm{R}^{1\times{c_J}}
    \label{eq:align-a2j}
\end{equation}
\vspace{-3mm}
\begin{equation}
    \textit{f}_{{M}\rightarrow{J}}(z_{M}): \mathbbm{R}^{1\times{c_M}} \rightarrow \mathbbm{R}^{1\times{c_J}}
    \label{eq:align-m2j}
\end{equation}
\vspace{-5mm}

\subsection{Learning Objectives}Proper design of learning objectives plays vital role in representation learning. After the audio embedding and motion embedding been aligned to same dimension, we discuss in the follow that there are two design options for learning objectives:\par
\vspace{-3mm}

\paragraph{Contrastive-Based Objectives}Contrastive-based loss is widely adopted in representation learning\cite{oord2018representation}\cite{radford2021learning}. For our case, given aligned motion and audio pairs, it aims to reduce the distance between their embedding or increase the similarity between them. On the contrary, for misaligned pairs, it attempts to increase the distance or reduce the similarity
between the embedding. During training, we construct mini batch samples containing $N$+1 pairs of motion and music, where 1 pair is stylistically aligned denoted as positive, and other $N$ pairs are misaligned denoted as negatives. We optimize the trainable MLPs ($\textit{f}_{{M}\rightarrow{A}}$, $\textit{f}_{{A}\rightarrow{M}}$, $\textit{f}_{{A}\rightarrow{J}}$, $\textit{f}_{{M}\rightarrow{J}}$) by minimizing the InfoNCE loss\cite{oord2018representation} as:
\vspace{-3mm}

\begin{equation}
    \mathcal{L}_{{M}\rightarrow{A}} = -\log{\frac{\exp{(z_{i}^{M} \cdot z_{i}^{A} / \tau)}}{\sum_{j=1}^{N}\exp{(z_{i}^{M} \cdot z_{j}^{A}) / \tau}}}
    \label{eq:contrastive-m2a}
\end{equation}

The final loss is: $\mathcal{L}_{{M}\leftrightarrow{A}}^{contr}=(\mathcal{L}_{{M}\rightarrow{A}}+\mathcal{L}_{{A}\rightarrow{M}})/2$.\par
\vspace{-3mm}

\paragraph{Clustering-Based Objectives}Unlike contrastive-based objectives, clustering-based objectives assumes relaxed pairwise alignment exists. We assume the stylistic representation of music and motion lay in joint high-dimension latent space. The embedding of motions or music of the same style are close to each other, while those with different style are apart from each other. Therefore, embeddings of motion and music of the same style form one latent subspace, or same cluster in latent space. As stated previously, the music-dance mapping follows a relaxed assumption. It is not necessary for the embedding of stylistic paired music and motion to be very close to each other in the latent space. Instead, their embedding should fall into the same subspace or cluster. Therefore, we don't construct positive and negative pairs for training. Instead, we attempt to group the embedding of aligned music and dance sequences into same cluster, while misaligned embedding should be clustered into different clusters. We assume music streams could be categorized into $\mathcal{C}^N$ genres as a prior. Therefore, for music embedding $z_{A}^{c_j}$ of genre $c_{j}$ and motion embedding $z_{M}^{c_j}$ corresponding to the same music style, we attempt to optimize the mapping $f_{{a}\rightarrow{b}}(z_{a})$ so that the mapped embedding $f_{{M}\rightarrow{b}}(z_{M})$ and $f_{{A}\rightarrow{b}}(z_{A})$ belong to same cluster $c_j$. Similarly, for music embedding $z_{A}^{c_j}$ and motion embedding $z_{M}^{c_k}$ belonging to different style $c_j$ and $c_k$, respectively, the mapped embedding are optimized to belong to different clusters. Let us define $K$ learnable embedding $\hat{c}_k$ representing cluster centers, we train the MLPs by optimizing the following objective:\par
\vspace{-3mm}

\begin{equation}
    \mathcal{L}_{intra}^{a} = \frac{1}{K}\sum_{i=1}^{K}{(1-\langle \tilde{z}_{a}^{c_i}, \hat{c}_i \rangle)}
    \label{eq:intra}
\end{equation}
\vspace{-3mm}
\begin{equation}
    \mathcal{L}_{inter}^{a} = \frac{1}{{K}({K-1})}\sum_{i=1}^{K}\sum_{j=1, {j}\neq{i}}^{K}{\langle \tilde{z}_{a}^{c_i}, \hat{c}_j \rangle}
    \label{eq:inter}
\end{equation}
\vspace{-3mm}
\begin{equation}
    \mathcal{L}_{reg} = \frac{1}{{K}({K-1})}\sum_{i=1}^{K}\sum_{j=1, {j}\neq{i}}^{K}{\langle \hat{c}_{i}, \hat{c}_j\rangle}
    \label{eq:reg}
\end{equation}
\vspace{-3mm}

where $\langle \cdot, \cdot \rangle$ is the similarity between two embeddings, $K$ is the number of styles, and $a$ denote either music or motion. The final loss is: $\mathcal{L}_{{M}\leftrightarrow{A}}^{cluster}=(\lambda_{1}\mathcal{L}_{intra}^{M}+\lambda_{2}\mathcal{L}_{inter}^{M}+\lambda_{3}\mathcal{L}_{intra}^{A}+\lambda_{4}\mathcal{L}_{inter}^{A}+\lambda_{5}\mathcal{L}_{reg}^{A})$.\par
\vspace{-3mm}

\paragraph{Classification Objectives}In addition, we adopt a classification loss as an auxiliary objective. We use a linear layer to project the mapped embedding in joint space to class probability distribution, and cross entropy loss is employed as supervision.\par

\begin{figure*}
    \centering
    \includegraphics[width=\textwidth]{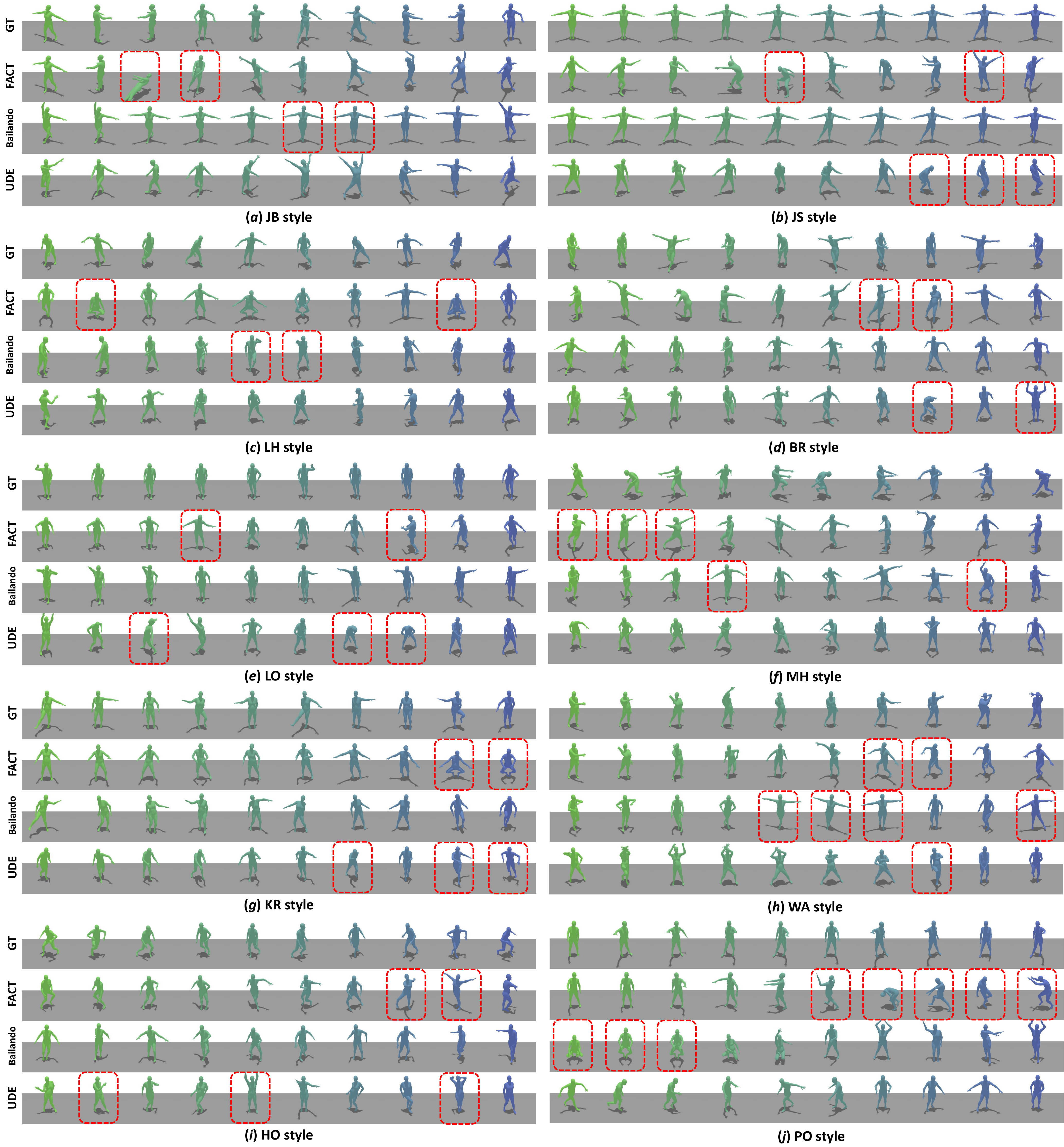}
    \caption{\textbf{Visual Comparison of Music-Dance Style Consistency.} We compare the generated motion sequences conditioned on different music styles with GTs. (a) \textbf{JB} means \textit{Ballet Jazz} style, (b) \textbf{JS} means \textit{Street Jazz} style, (c) \textbf{LH} means \textit{LA Hiphop} style, (d) \textbf{BR} means \textit{Break} style, (e) \textbf{LO} means \textit{Lock} style, (f) \textbf{MH} means \textit{Middle Hiphop} style, (g) \textbf{KR} means \textit{Krump} style, (h) \textbf{WA} means \textit{Waacking} style, (i) \textbf{HO} means \textit{House} style, (j) \textbf{PO} means \textit{Pop} style. For each dance, we adopt 10sec motion segment and evenly sample 10 frames. The $\color{red}{\rm dashed}$ $\color{red}{\rm box}$ indicates poses that are style inconsistent with GTs.}
    \label{fig:comparison}
\end{figure*}

\begin{table*}[]
    \centering
    \resizebox{1.0\linewidth}{!}{
        \begin{tabular}{cc|cccccc|cccccc|c}
            \hline
            \multicolumn{2}{c|}{Method} & \multicolumn{6}{c|}{Music} & \multicolumn{6}{c|}{Motion} & \multirow{2}{*}{Simi. $\uparrow$ } \\
            \cline{1-14}
             Alignment & Objective & Acc. $\uparrow$ & Top-1 Retr. $\uparrow$ & Top-3 Retr. $\uparrow$ & Intra. $\downarrow$ & Inter. $\uparrow$ & I2I. $\downarrow$ & Acc. $\uparrow$ & Top-1 Retr. $\uparrow$ & Top-3 Retr. $\uparrow$ & Intra. $\downarrow$ & Inter. $\uparrow$ & I2I. $\downarrow$ \\
             \hline
             $f_{{M}\rightarrow{A}}$ & $\mathcal{L}_{{M}\leftrightarrow{A}}^{cluster}$ & 44.00\% & 32.80\% & 45.60\% & 1.26 & 1.38 & 0.91 & 93.80\% & \cellcolor{blue!15}{94.00}\% & 99.60\% & 0.24 & 1.39& 0.18 & 0.21 \\
             $f_{{A}\rightarrow{M}}$ & $\mathcal{L}_{{M}\leftrightarrow{A}}^{cluster}$ & 59.20\% & \cellcolor{blue!15}{58.40}\% & 63.20\% & \cellcolor{blue!15}{0.77} & 1.33& 0.60 & 91.80\% & 57.20\% & 85.00\% & 1.25 & 1.41 & 0.89 & 0.18 \\
             $f_{{M}\rightarrow{J}}$+$f_{{A}\rightarrow{J}}$ & $\mathcal{L}_{{M}\leftrightarrow{A}}^{contr}$ & \cellcolor{blue!15}{60.80}\% & - & - & - & - & - & 92.20\% & - & - & - & - & - & 0.44 \\
             $f_{{M}\rightarrow{J}}$+$f_{{A}\rightarrow{J}}$ & $\mathcal{L}_{{M}\leftrightarrow{A}}^{cluster}$ & 57.60\% & \cellcolor{blue!15}{58.40\%} & \cellcolor{blue!15}{76.00}\% & 0.83 & \cellcolor{blue!15}{1.43} & \cellcolor{blue!15}{0.59} & \cellcolor{blue!15}{94.20}\% & 93.40\% & \cellcolor{blue!15}{99.80}\% & \cellcolor{blue!15}{0.24} & \cellcolor{blue!15}{1.48} & \cellcolor{blue!15}{0.16} & \cellcolor{blue!15}{0.47} \\
             \hline
            \hline
            % \multicolumn{2}{c|}{Method} & \multicolumn{6}{c|}{Music} & \multicolumn{6}{c|}{Motion} & \multirow{2}{*}{Simi. $\uparrow$ } \\
            % \cline{1-14}
            %  Alignment & Objective & Acc. $\uparrow$ & Top-1 Retr. $\uparrow$ & Top-3 Retr. $\uparrow$ & Intra. $\downarrow$ & Inter. $\uparrow$ & I2I. $\downarrow$ & Acc. $\uparrow$ & Top-1 Retr. $\uparrow$ & Top-3 Retr. $\uparrow$ & Intra. $\downarrow$ & Inter. $\uparrow$ & I2I. $\downarrow$ \\
            %  \hline
             $f_{{M}\rightarrow{A}}$ & $\mathcal{L}_{{M}\leftrightarrow{A}}^{cluster}$ & 52.21\% & 32.71\% & 82.23\% & 1.05 & 1.24 & 0.85 & 55.16\% & 58.17\% & 86.72\% & 0.89 & 1.35& 0.67 & 0.32 \\
             $f_{{A}\rightarrow{M}}$ & $\mathcal{L}_{{M}\leftrightarrow{A}}^{cluster}$ & 75.17\% & 75.94\% & \cellcolor{blue!15}{93.71}\% & 0.60 & \cellcolor{blue!15}{1.37}& 0.45 & 38.55\% & 16.29\% & 75.24\% & 1.24 & 1.30 & 0.95 & 0.22 \\
             $f_{{M}\rightarrow{J}}$+$f_{{A}\rightarrow{J}}$ & $\mathcal{L}_{{M}\leftrightarrow{A}}^{contr}$ & 74.02\% & - & - & - & - & - & 57.15\% & - & - & - & - & - & \cellcolor{blue!15}{0.62} \\
             $f_{{M}\rightarrow{J}}$+$f_{{A}\rightarrow{J}}$ & $\mathcal{L}_{{M}\leftrightarrow{A}}^{cluster}$ & \cellcolor{blue!15}{77.04}\% & \cellcolor{blue!15}{77.48\%} & 92.75\% & \cellcolor{blue!15}{0.51} & 1.35 & \cellcolor{blue!15}{0.39} & \cellcolor{blue!15}{58.88}\% & \cellcolor{blue!15}{60.48}\% & \cellcolor{blue!15}{88.90}\% & \cellcolor{blue!15}{0.81} & \cellcolor{blue!15}{1.36} & \cellcolor{blue!15}{0.61} & 0.59 \\
             \hline
        \end{tabular}
    }
    \caption{\textbf{Evaluation Results on Music Streams and GT Motion Sequences.} We report the quantitative results of the variants of our method in music-dance style consistency evaluation on the test set of AIST++\cite{li2021ai} and AIOZ-GDANCE\cite{le2023music}. We measure the style estimation accuracy, style retrieval accuracy, as well as style consistency score on both music streams and dance motion sequences. $\colorbox{blue!15}{\rm Indicate best results}$.}
    \label{tab:gt}
\end{table*}

\begin{table}[]
    \centering
    \resizebox{1.0\linewidth}{!}{
        \begin{tabular}{c|ccccccc}
            \hline
            Method & Acc. $\uparrow$ & Top-1 Retr. $\uparrow$ & Top-3 Retr. $\uparrow$ & Intra. $\downarrow$ & Inter. $\uparrow$ & I2I. $\downarrow$ & Simi. $\uparrow$ \\
            \hline
            FACT\cite{li2021ai} & 17.67\% & 17.96\% & 43.47\% & 1.25 & 1.38 & 0.92 & -0.03 \\
            Bailando\cite{siyao2022bailando} & \cellcolor{blue!15}{37.04\%} & \cellcolor{blue!15}{38.01\%} & \cellcolor{blue!15}{60.23\%} & \cellcolor{blue!15}{1.02} & \cellcolor{blue!15}{1.41} & \cellcolor{blue!15}{0.74} & \cellcolor{blue!15}{0.32} \\
            UDE\cite{zhou2023ude} & 20.69\% & 19.84\% & 40.10\% & 1.22 & 1.38 & 0.89 & 0.17 \\
            \hline
        \end{tabular}
    }
    \caption{\textbf{Evaluation Results on Generated Motion Sequences.} We conduct evaluation on the generated motion sequences of different methods \cite{li2021ai}\cite{siyao2022bailando}\cite{zhou2023ude}. $\colorbox{blue!15}{\rm Indicate best results}$.}
    \label{tab:pred}
\end{table}

% \begin{table}[]
%     \centering
%     \resizebox{1.0\linewidth}{!}{
%         \begin{tabular}{c|cc|cccc}
%             \hline
%             \multirow{2}{*}{Method} & \multicolumn{2}{c|}{MDSC} & \multicolumn{4}{c}{Kinetic + Manual} \\
%             \cline{2-7}
%             & FID $\downarrow$ & Div $\rightarrow$ & FID\(_k\) $\downarrow$ & FID\(_m\) $\downarrow$ & Div\(_k\) $\rightarrow$ & Div\(_m\) $\rightarrow$ \\
%             \hline
%             GT & 0.09 & 1.33 & 17.10 & 10.60 & 8.19 & 7.45 \\
%             \hline
%             FACT\cite{li2021ai} & 0.85 & 0.78 & 35.35 & 22.11 & 5.94 & 6.18 \\
%             Bailando\cite{siyao2022bailando} & 0.15 & 1.31 & 28.16 & 9.62 & \cellcolor{blue!15}{7.83} & \cellcolor{blue!15}{6.34} \\
%             UDE\cite{zhou2023ude} & \cellcolor{blue!15}{0.11} & \cellcolor{blue!15}{1.32} & \cellcolor{blue!15}{17.25} & \cellcolor{blue!15}{8.69} & 7.78 & 5.81 \\
%             \hline
%         \end{tabular}
%     }
%     \caption{\textbf{Comparison of FID and Div.} We also compute FID and Div on GT and \cite{li2021ai}\cite{siyao2022bailando}\cite{zhou2023ude} using the embedding obtained by our method, and compare the relative relationship among them with the results reported in \cite{zhou2023ude}. $\colorbox{blue!15}{\rm Indicate best results}$.}
%     \label{tab:fid}
% \end{table}

\begin{figure}[h]
    \centering
    \includegraphics[width=0.9\linewidth]{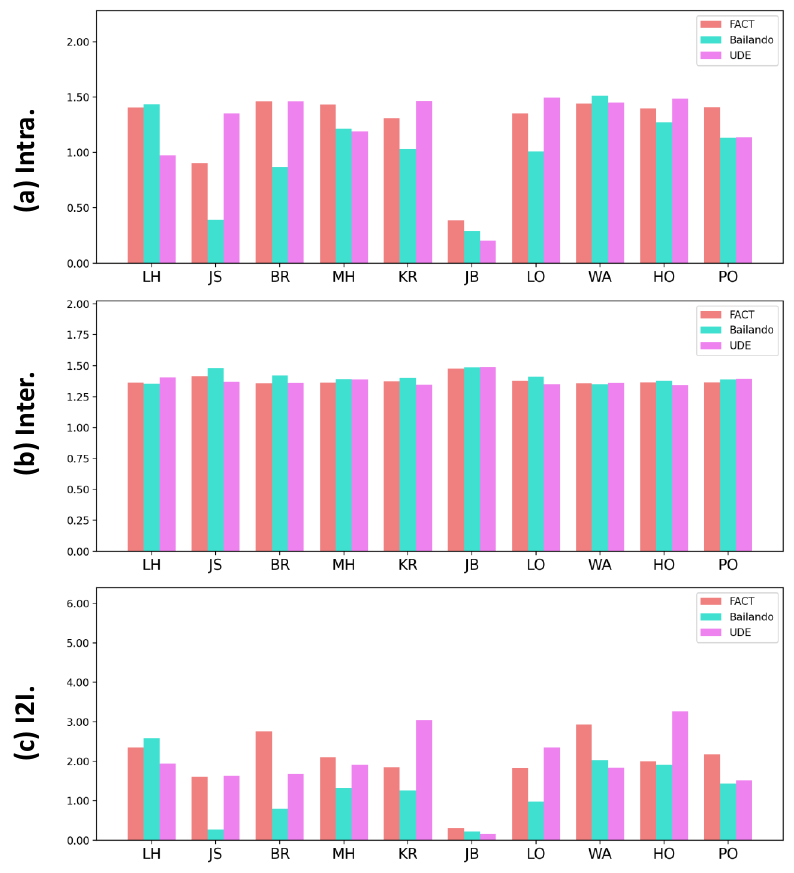}
    \caption{\textbf{Results of Cluster Distance of SOTAs.} We calculate the intracluster(\textit{\textbf{Intra.}}) distance, intercluster distance(\textit{\textbf{Inter.}}), and intra-to-inter(\textit{\textbf{I2I.}}) between generated motion embedding and learned cluster centers embedding and report their mean and variance.}
    \label{fig:stats}
\end{figure}

\section{Experiments}We evaluate our method on a widely adopted public available dataset AIST++\cite{li2021ai} and AIOZ-GDANCE\cite{le2023music}. We conduct thorough quantitative and qualitative analysis on several music-driven motion generation methods\cite{li2021ai}\cite{siyao2022bailando}\cite{zhou2023ude} to validate that our method is an appropriate design for music-dance style consistency assessment. We also build a benchmark(Tab. \ref{tab:pred}) using our method for future research.\par

\subsection{Implementation Details}

\paragraph{Data Preprocessing}For motion sequence, we adopt the SMPL representation\cite{SMPL:2015}. We represent each pose frame as a 75D vector, where the first 3D are the root trajectory, the 3-6D are the root orientation in rotation vector format, and the rest 69D are the rotation vectors of each joints relative to their parents. For music sequence, we read the raw acoustic waveform data and sample to 16KHz. The window size of each training sequence is 160. We follow the standard rule to split the data into training and validation set.\par
\vspace{-3mm}

\paragraph{Motion Encoder}For motion auto-encoder, we adopt transformer encoder as the encoder architecture, and transformer decoder as the decoder architecture. For both encoder and decoder, the number of layers are 6, the number of attention heads are 4, and the hidden dimension size is 768. We train the auto-encoder on AMASS\cite{AMASS:ICCV:2019}, AIST++\cite{li2021ai} and AIOZ-GDANCE\cite{le2023music}.\par
\vspace{-3mm}

\paragraph{Music-Dance Style Alignment}For the mapping layer, we adopt a 2-layer MLP and project the input embedding to 256D embedding.\par

\subsection{Evaluation Metrics}To evaluate the effectiveness of our method, we define following metrics: 1) \textit{Style Classification Accuracy}(\textbf{Acc.}). We estimate the style class of music and motion embeddings in joint space with one logits head. 2) \textit{Style Retrieval}(\textbf{Retr.}). We assume a good style evaluation model is able to map input sequences to embedding which are well clustered in latent space. Consequently, for either music streams or motion sequences, we encode them, and calculate the distance between their embedding and each cluster centers embedding. If the distance to correct center embedding ranks No. \textit{k} closest, it is considered as Top-\textit{k} Retrieval accuracy. 3) \textit{Intra-Cluster Distance}(\textbf{Intra.}). We measure the distance between input embedding to the correct center embedding. It is expected the more stylistically consistent the input sequence is, the closer the embedding distance is. 4) \textit{Inter-Cluster Distance}(\textbf{Inter.}). On the contrary to \textit{Intra.}, we expect the distance between input sequence embedding and incorrect center embedding as large as possible. This metric measures how stylistically inconsistent the input are to misaligned styles. 5) \textit{Intra-2-Inter}(\textbf{I2I.}). This metric correlates \textit{Inter.} and \textit{Intra.} and calculated as: $I2I = \frac{Intra.}{Inter.}$ and uses a scalar to measure how consistent an input sequence is to the target style. The smaller the $I2I$ is, the more stylistically consistent the input sequence is. 6) \textit{Embedding Similarity}(\textbf{Simi.}). This metric measures the cosine similarity between audio embedding $z_{A}$ and motion embedding $z_{M}$. Higher score indicates higher style consistency.\par

\begin{table*}
    \centering
    \resizebox{1.0\linewidth}{!}{
        \begin{tabular}{ccc|cccccc|cccccc|c}
            \hline
            \multicolumn{3}{c|}{Losses} & \multicolumn{6}{c|}{Music} & \multicolumn{6}{c|}{Motion} & \multirow{2}{*}{Simi. $\uparrow$} \\
            \cline{1-15}
            $\mathcal{L}_{intra}$ & $\mathcal{L}_{inter}$ & $\mathcal{L}_{reg}$ & Acc. $\uparrow$ & Top-1 Retr. $\uparrow$ & Top-3 Retr. $\uparrow$ & Intra. $\downarrow$ & Inter. $\uparrow$ & I2I. $\downarrow$ & Acc. $\uparrow$ & Top-1 Retr. $\uparrow$ & Top-3 Retr. $\uparrow$ & Intra. $\downarrow$ & Inter. $\uparrow$ & I2I. $\downarrow$ \\
            \hline
            \checkmark & & & 45.60\% & 13.60\% & 27.20\% & 1.42 & 1.41 & 1.00 & 91.00\% & 0.40\% & 10.40\% & 1.42 & 1.42 & 1.00 & 0.11 \\
            \checkmark & \checkmark & & \cellcolor{blue!15}{59.20\%} & \cellcolor{blue!15}{59.20\%} & 66.40\% & \cellcolor{blue!15}{0.75} & 1.32& \cellcolor{blue!15}{0.59} & 92.80\% & 92.40\% & 99.60\% & \cellcolor{blue!15}{0.23} & 1.39& 0.17 & 0.42 \\
            \checkmark & \checkmark & \checkmark & 57.60\% & 58.40\% & \cellcolor{blue!15}{76.00\%} & 0.83 & \cellcolor{blue!15}{1.43} & \cellcolor{blue!15}{0.59} & \cellcolor{blue!15}{94.20\%} & \cellcolor{blue!15}{93.40\%} & \cellcolor{blue!15}{99.80\%} & 0.24 & \cellcolor{blue!15}{1.48} & \cellcolor{blue!15}{0.16} & \cellcolor{blue!15}{0.47} \\
            \hline
            % \hline
            % \multicolumn{3}{c|}{Losses} & \multicolumn{6}{c|}{Music} & \multicolumn{6}{c|}{Motion} & \multirow{2}{*}{Simi. $\uparrow$} \\
            % \cline{1-15}
            % $\mathcal{L}_{intra}$ & $\mathcal{L}_{inter}$ & $\mathcal{L}_{reg}$ & Acc. $\uparrow$ & Top-1 Retr. $\uparrow$ & Top-3 Retr. $\uparrow$ & Intra. $\downarrow$ & Inter. $\uparrow$ & I2I. $\downarrow$ & Acc. $\uparrow$ & Top-1 Retr. $\uparrow$ & Top-3 Retr. $\uparrow$ & Intra. $\downarrow$ & Inter. $\uparrow$ & I2I. $\downarrow$ \\
            % \hline
            % \checkmark & & & 45.60\% & 13.60\% & 27.20\% & 1.42 & 1.41 & 1.00 & 91.00\% & 0.40\% & 10.40\% & 1.42 & 1.42 & 1.00 & 0.11 \\
            % \checkmark & \checkmark & & \cellcolor{blue!15}{59.20\%} & \cellcolor{blue!15}{59.20\%} & 66.40\% & \cellcolor{blue!15}{0.75} & 1.32& \cellcolor{blue!15}{0.59} & 92.80\% & 92.40\% & 99.60\% & \cellcolor{blue!15}{0.23} & 1.39& 0.17 & 0.42 \\
            % \checkmark & \checkmark & \checkmark & 57.60\% & 58.40\% & \cellcolor{blue!15}{76.00\%} & 0.83 & \cellcolor{blue!15}{1.43} & \cellcolor{blue!15}{0.59} & \cellcolor{blue!15}{94.20\%} & \cellcolor{blue!15}{93.40\%} & \cellcolor{blue!15}{99.80\%} & 0.24 & \cellcolor{blue!15}{1.48} & \cellcolor{blue!15}{0.16} & \cellcolor{blue!15}{0.47} \\
            % \hline
        \end{tabular}
    }
    \caption{\textbf{Ablation on loss terms} We train our method using design option $f_{{M}\rightarrow{J}}$+$f_{{A}\rightarrow{J}}$ with learning objective $\mathcal{L}^{cluster}$, and compare the quantitative results between using training with 1) $\mathcal{L}_{intra}$ only, 2) $\mathcal{L}_{intra}$+$\mathcal{L}_{inter}$, and 3) full clustering-based loss as $\mathcal{L}_{intra}$+$\mathcal{L}_{inter}$+$\mathcal{L}_{reg}$. $\colorbox{blue!15}{\rm Indicate best results}$.}
    \label{tab:ablation}
\end{table*}

\begin{table}
    \centering
    \resizebox{0.85\linewidth}{!}{
        \begin{tabular}{c|ccc}
            \hline
            Method & Music Acc. $\uparrow$ & Motion Acc. $\uparrow$ & Simi. $\uparrow$ \\
            % \cline{2-4}
            \hline
            w/o $\hat{c}$ & 53.40\% & 92.20\% & 0.46 \\
            w/ $\hat{c}$ & \cellcolor{blue!15}{57.60\%} & \cellcolor{blue!15}{94.20\%}  & \cellcolor{blue!15}{0.47} \\
            \hline
        \end{tabular}
    }
    \caption{\textbf{Ablation on Learning Strategy} We train our method using design option $f_{{M}\rightarrow{J}}$+$f_{{A}\rightarrow{J}}$ with different learning strategies. 1) w/o $\hat{c}$ means we train the model without learnable cluster center embedding $\hat{c}$, it is assumed the number of cluster is unknown. 2) w/ $\hat{c}$ assumes the number of cluster is known. $\colorbox{blue!15}{\rm Indicate best results}$.}
    \label{tab:ablation-2}
\end{table}

\subsection{Results}

\paragraph{Evaluation Results on GT}We apply our method with three different variant of designs on the test set of AIST++\cite{li2021ai} and AIOZ-GDANCE\cite{le2023music}.The quantitative results are reported at Tab. \ref{tab:gt}. As we can see, if we choose to align motion embedding $z_{M}$ to audio embedding $z_{A}$, we found that the model is not able to understand the music's style. This is because the assumption that the pretrained music encoder is able to extract semantically righ information for style understanding does not hold. Similarly, if we choose to map audio embedding $z_{A}$ to align with motion embedding $z_{M}$, it is shown that the motion style is not understood well. This is also because the assumption that pretrained motion encoder captures style representative embedding dose not hold neither. The design of mapping audio embedding $z_{A}$ and motion embedding $z_{M}$ to joint embedding space, however, shows that both music and dance motion style prediction, retrieval, and consistency evaluation achieve high score. This indicates that jointly learning the mappings $f_{{M}\rightarrow{J}}$ and $f_{{A}\rightarrow{J}}$ is able to align $z_{M}$ and $z_{A}$ in a representative informative latent space, and outperforms the other two variants cross-modality alignment. 
% In terms of the impact of learning objectives, we found that contrastive-based objective is slightly better at estimating music style, while underperforms at estimating motion style and measuring cross-modality embedding similarity. Conclusion could be drawn that aligning both embedding to joint space as $f_{{M}\rightarrow{J}}$ and $f_{{A}\rightarrow{J}}$, and training with objective $\mathcal{L}^{cluster}$ is an optimal design choice.
\par 
\vspace{-3mm}

\paragraph{Evaluation Results on Generated Motions}We apply our method on the generated dance motion sequences of three different methods \cite{li2021ai}\cite{siyao2022bailando}\cite{zhou2023ude}, and conduct both quantitative and visual analysis to validate that our method is able to evaluate the music-dance style consistency. Fig. \ref{fig:stats} shows the intra-cluster distances, inter-cluster diatances, and intra-to-inter ratios of generated motion of each method, organized by music styles. The visual results of generated motions by each method and the GTs are selected and shown in Fig. \ref{fig:comparison}. For each dance sequence, we randomly crop a 10$\textit{sec}$ segment with FPS=30, which is in total 300 frames, and evenly sample 10 frames for visualization. We use red dashed box to indicate the poses that are style inconsistent with the driving music.\par

\setlength{\parindent}{1em}For example, the intra-to-inter ratio($\textbf{I2I.}$) of $\textbf{JB}$ style indicates that the MDSC(music-dance style consistency) among three methods follows UDE\cite{zhou2023ude} $>$ Bailando\cite{siyao2022bailando} $>$ FACT\cite{li2021ai}, but all method generate highly consistent results. Correspondingly, Fig. \ref{fig:comparison}(a) shows that although few poses are identified with artifacts and style inconsistency, most of the poses are style consistent. Another example is $\textbf{JS}$ style. In Fig. \ref{fig:stats}(c), the $\textit{\textbf{I2I.}}$ of $\textbf{JS}$ shows the style consistency among three methods are Bailando\cite{siyao2022bailando} $>$ FACT\cite{li2021ai} $>$ UDE\cite{zhou2023ude}, and among three, the consistency of Bailando\cite{siyao2022bailando} is much higher than other two methods, and the consistency of UDE\cite{zhou2023ude} is relatively low. This conclusion could be drawn from visualization analysis in Fig. \ref{fig:comparison}(b) as well. As we can see in Fig. \ref{fig:comparison}(b), the dance style of $\textbf{JS}$ music is stretching out both arms out in a T-pose and spreading legs gently. For FACT\cite{li2021ai}, artifacts and inconsistent poses are identified by red dashed boxes, the inconsistent pose looks like a $\textbf{JB}$ style rather than $\textbf{JS}$ style. The dance motion of UDE\cite{zhou2023ude} presents more inconsistent poses. As shown in the figure, the last 3 poses present swing style, which is obviously inconsistent with desired $\textbf{JS}$ style. These comparisons show that our method serves as a better metric in assessing the music-motion consistency in terms of music and dance style.\par

%%%%%%%%%% by Zhou Zixiang
The evaluation results in Tab. \ref{tab:pred} shows discrepancy with the user study reported in \cite{zhou2023ude}, in which the generated dances of \cite{zhou2023ude} are preferred than \cite{siyao2022bailando, li2021ai} by the participants. We argue that this is due to different evaluation focuses. In \cite{zhou2023ude}, participants are expected to pay more attention to the motion quality, specifically, whether the motion is smooth and natural. However, the primary concern in Tab. \ref{tab:pred} is the style consistency between dances and musics. This also justifies the necessity of proposing style consistency metric.\par

\vspace{-3mm}

% \paragraph{Evaluation Results on Fidelity and Diversity} We also show that our method is able to assess fidelity and diversity as well. As our method maps motion seuqence into high dimension embedding, we show that these embedding could be used as indicator of fidelity(\textbf{FID}) and diversity(\textbf{Div}) assessment. We compute the \textbf{FID} and \textbf{Div} on GT and generated motion of \cite{li2021ai}\cite{siyao2022bailando}\cite{zhou2023ude}, and compare the relative relationship among them with the results reported in \cite{zhou2023ude} using different method. As shown in Tab. \ref{tab:fid}, we can observe same relative relationship on \textbf{FID} using both methods. For \textbf{Div}, although ours evaluation suggest \cite{zhou2023ude} achieves best diversity, while other method's result suggest \cite{siyao2022bailando} has best diversity, their difference are slight.\par
% \vspace{-3mm}

\subsection{Ablation}We conduct ablation study to explore the effectiveness of terms of $\mathcal{L}_{intra}$, $\mathcal{L}_{inter}$, and $\mathcal{L}_{reg}$. We report the quantitative comparison in Tab. \ref{tab:ablation}. We design three experiments with different combination of clustering terms to validate their effectiveness, and all experiments adopt same setting: $f_{{M}\rightarrow{J}}+f_{{A}\rightarrow{J}}$, $\mathcal{L}^{cluster}$. The first experiment uses only $\mathcal{L}_{intra}$ term, the second experiment adopts $\mathcal{L}_{intra}+\mathcal{L}_{inter}$, and the third one takes full $\mathcal{L}^{cluster}$ as its training objective. As we can see, the model trained with only $\mathcal{L}_{intra}$ does not learn representative encoding capability. Its estimation on music style is comparatively lower and the style retrieval accuracy for both music and motion are largely lower than other experiments. In addition, It computes worst $\textit{I2I}$ score and $\textit{Simi}$ score. The term $\mathcal{L}_{inter}$ affects the model's capability a lot. As we can see, the model training with $\mathcal{L}_{intra}+\mathcal{L}_{inter}$ achieves much better capability in estimating style, retrieving correct style from embedding, and clustering an input sequence to correct style cluster. The impact of $\mathcal{L}_{reg}$ is not as large as $\mathcal{L}_{inter}$, but it still improves the performance of the model.\par

\setlength{\parindent}{1em}We also ablate to explore the effectiveness of learning strategy. As we adopt clustering-based objective, there is two alternatives. 1) The number of clusters is unknown, and 2) the number of clusters is known. For option 2), cluster centers $\hat{c}$ are learned jointly to facilitate the learning of our method. We evaluate the methods trained with two strategies on AIST++\cite{li2021ai} and report the results in Tab. \ref{tab:ablation-2}. As we can see, learning without knowing the number of clusters performs worse.\par
\section{Conclusion}We propose \textbf{MDSC}, the first method in measuring music-motion style consistency, in this paper. We adopt pretrained encoders for music embedding and motion embedding, and adopt MLPs to align them in joint latent space. We learn the mapping using clustering-based objective instead of constrastive-based objective. We conduct thorough experiments to validate that our method is able to assess the music-dance style consistency, and we provide benchmarks in Tab. \ref{tab:pred} on three different music-driven methods. 
{
    \small
    \bibliographystyle{ieeenat_fullname}
    \bibliography{main}
}

% WARNING: do not forget to delete the supplementary pages from your submission 

\clearpage
\appendix

\twocolumn[
\begin{@twocolumnfalse}
\section*{\centering{Supplementary Material}}
\end{@twocolumnfalse}
]

%----------------------------------------------------------------------
\section{Visualization of Multimodality Alignment} 
\begin{figure*}
    \centering
    \includegraphics[width=\textwidth]{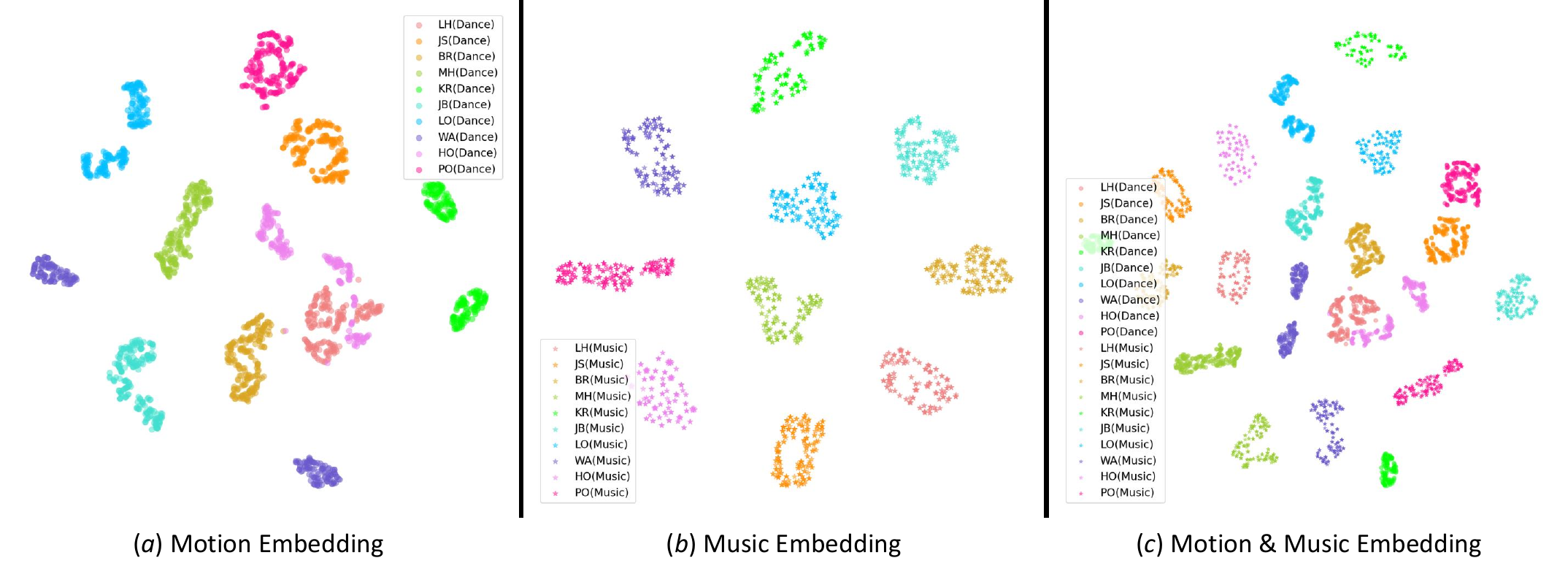}
    \caption{\textbf{Visualization of Embeddings Encoded by Pre-Trained Encoders.} We show why embeddings obtained directly from pre-trained encoders are not aligned. a) Shows the results of motion embeddings obtained by $\mathcal{E}_{M}$, and embeddings belonging to different styles are colored accordingly. b) Shows the results of music embeddings encoded by $\mathcal{E}_{A}$. Similarly, different color indicates differnt music styles. c) Shows the motion and music embeddings are not aligned in joint space. $\ast$ indicates music embedding, and $\bullet$ indicates motion embedding.}
    \label{fig:supp-embed}
\end{figure*}

We show that our pretrained motion encoder $\mathcal{E}_{M}$ and music encoder $\mathcal{E}_{A}$ are able to encode input into embeddings in latent space which are easy to be clustered. However, as stated previously, they are not necessarily aligned in latent space because the encoders are trained separately. As shown in Fig. \ref{fig:supp-embed}(a), (b), the embedding of motion sequence and music belonging to different styles are colored in differently, and it is clearly observed that embedding belonging to same style are close together, forming clusters in latent space. However, the embeddings obtained by pretrained encoders are not aligned in latent space. As shown in Fig. \ref{fig:supp-embed}(c), motion embedding and music embedding belonging to same style locate separately, indicating large distance in high dimension latent space.\par
\vspace{3mm}

\begin{figure*}
    \centering
    \includegraphics[width=\textwidth]{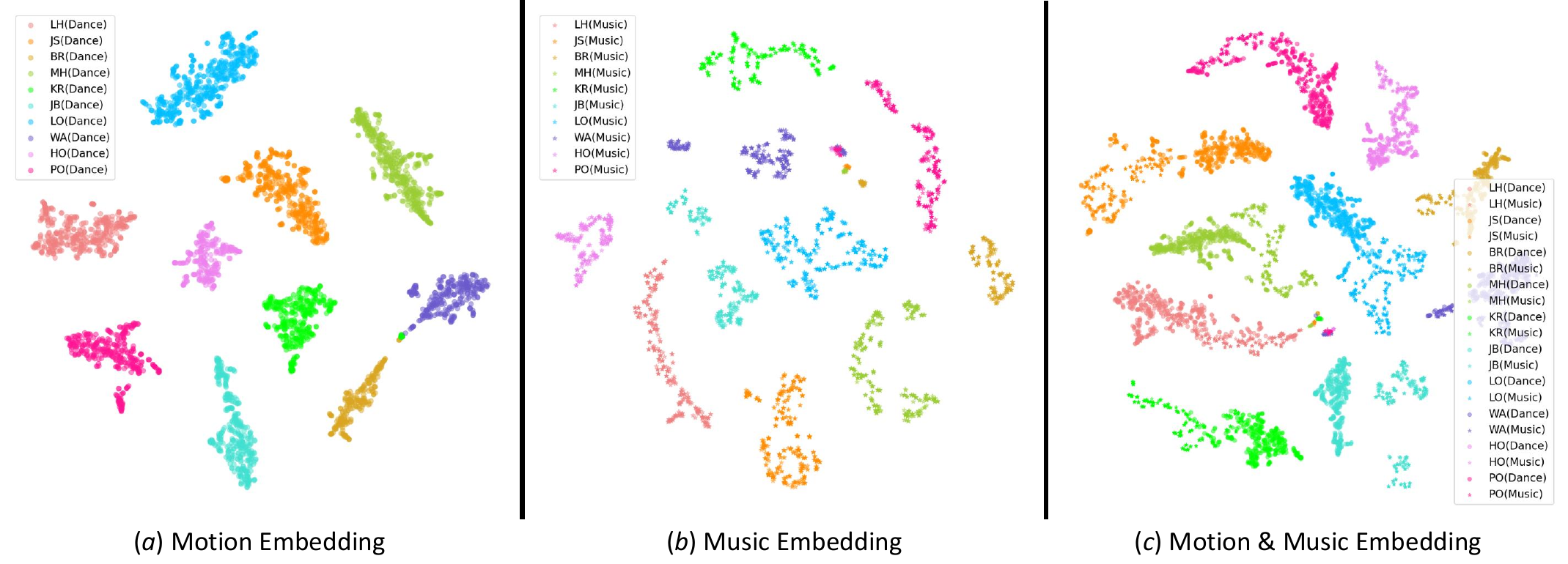}
    \caption{\textbf{Visualization of Embeddings after Projection in Joint Space.} We show the results after projected by adopting MLPs($f_{M \rightarrow J}$, $f_{A \rightarrow J}$) to joint latent space. a) Shows the motion embeddings in joint latent space. Embeddings belonging to different styles are colored differently. b) Music embeddings in joint space are visualized in similar manner. c) The motion embedding and music embedding are projected and aligned in joint latent space. $\ast$ indicates music embedding, and $\bullet$ indicates motion embedding.}
    \label{fig:supp-embed2}
\end{figure*}

\setlength{\parindent}{1em} We show in Fig. \ref{fig:supp-embed2} that adopting mappings $f_{M \rightarrow J}$ and $f_{A \rightarrow J}$ helps in aligning the motion and music embeddings in latent space remarkably. Fig. \ref{fig:supp-embed2}(a), (b) clearly conveys the mapping $f_{M \rightarrow J}$ and $f_{A \rightarrow J}$ preserve the manifold of embedding in high dimension space. In the meantime, Fig. \ref{fig:supp-embed2}(c) demonstrates the motion and music embeddings are well aligned in latent space.

%----------------------------------------------------------------------
\section{Evaluation Results on Synthesis Videos} 
As shown in the supplementary video, we conducted a demonstration to showcase the accurate evaluation of dance and music consistency by our model. To achieve this, we synthesized several videos. Our approach involved sampling dance-music pairs from the AIOZ-GDANCE\cite{le2023music} dataset. In each video, there are two ground truth group dances that aligned with the music style, as well as two randomly selected dance movements from other styles. The Fig. \ref{fig:Simi} below illustrates the capability of our evaluation method in effectively discerning between matching dance-music pairs and inconsistent ones.

\begin{figure}[t]
    \centering
    \includegraphics[width=1\linewidth]{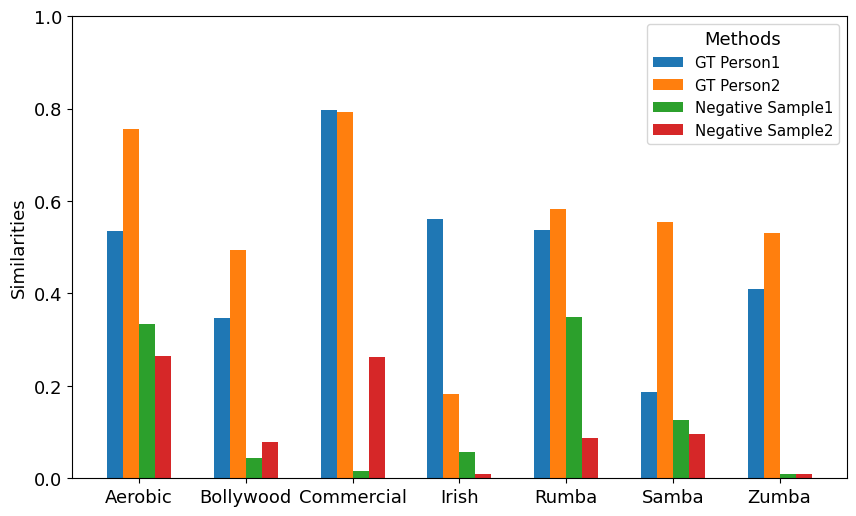}
    \caption{Evaluation results of our methods on synthesis videos of different styles in AIOZ-GDANCE\cite{le2023music} dataset. Dance-music pairs sampled from datasets achieve higher similarities compared to the negative samples. The similarities of negative sample of Irish and Zumba style are negative, so they are set to 0.01 for aesthetic purposes.}
    \label{fig:Simi}
\end{figure}

\begin{figure}[t]
    \centering
    \includegraphics[width=0.8\linewidth]{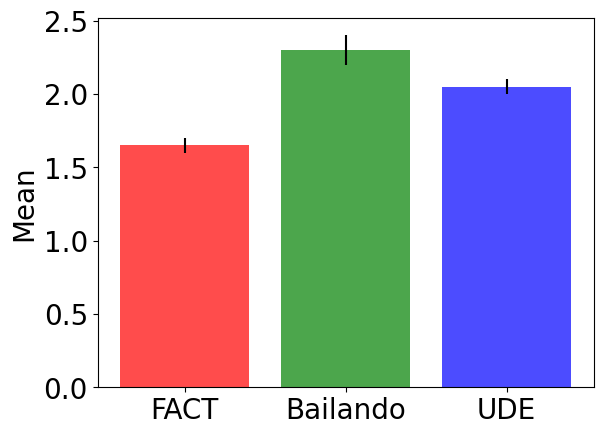}
    \caption{Average score of different algorithms annotated by 3 independent users. The results of user study are consistent with our metrics as shown in Sec. $\color{red}{4.3}$.}
    \label{fig:user_study}
\end{figure}

%----------------------------------------------------------------------
\section{More Evaluation Results on Generated Motions}
\subsection{Visualization of Generated Motions}
We present more dance motions generated by three different methods\cite{li2021ai}\cite{siyao2022bailando}\cite{zhou2023ude}, and the corresponding ground truth dances in the supplementary video and the following figures. We organize the results by the style of the conditioning music. The generated dances and the GTs are sampled at FPS=30. For each music style, we randomly select 6 groups of the results from GT and \cite{li2021ai}\cite{siyao2022bailando}\cite{zhou2023ude}. respectively, and we crop the generated dance motions to segments of 30 \textit{sec} length, and finally we evenly sample 10 frames per dance for the visualization. As shown in Sec. $\color{red}{4.3}$, we report the average MDSC of three methods organized by styles. It is recommended to go through the visualization and the results of Fig. $\color{red}{4}$ at the same time. For convenience, we list the relative relationship of MDSC of three methods\cite{li2021ai}\cite{siyao2022bailando}\cite{zhou2023ude} below:\par

\begin{itemize}
    \item[$\bullet$] $\textbf{BR}$ style, Bailando $>$ UDE $\approx$ FACT. 
    \item[$\bullet$] $\textbf{HO}$ style, Bailando $>$ FACT $>$ UDE.
    \item[$\bullet$] $\textbf{JB}$ style, Bailando $>$ FACT $>$ UDE.
    \item[$\bullet$] $\textbf{JS}$ style, Bailando $>$ FACT $>$ UDE.
    \item[$\bullet$] $\textbf{KR}$ style, Bailando $>$ FACT $>$ UDE.
    \item[$\bullet$] $\textbf{LH}$ style, UDE $>$ FACT $\approx$ Bailando.
    \item[$\bullet$] $\textbf{LO}$ style, Bailando $>$ FACT $>$ UDE.
    \item[$\bullet$] $\textbf{MH}$ style, UDE $>$ Bailando $>$ FACT.
    \item[$\bullet$] $\textbf{PO}$ style, UDE $\approx$ Bailando $>$ FACT.
    \item[$\bullet$] $\textbf{WA}$ style, FACT $\approx$ UDE $>$ Bailando.
\end{itemize}

\subsection{User Study}
Besides, to evaluate human preference of style consistency between music and dance. We apply user study on the generated motions of different music-conditioned dance generation algorithms. For each algorithm, we generate thirty 6-second clips. Five independent users are invited to rate the clips with score ranging from 1 to 3. $10$ ground-truth clips are first represented before rating. The average score with error bar is shown in Fig.~\ref{fig:user_study}. The results of user study are consistent with our metrics as shown in Sec. $\color{red}{4.3}$.

% aistpp 10 style
% \begin{itemize} 
%    \item GT 
%    \item Negative Sample 
%    \item Positive Sample
%    \item Bailando
%    \item UDE
%    \item FACT
% \end{itemize}

% gdance 7 style
% \begin{itemize} 
%    \item GT 
%    \item Negative Sample 
%    \item Positive Sample among different dancer
% \end{itemize}

\begin{figure*}
    \centering
    \includegraphics[width=0.88\textwidth]{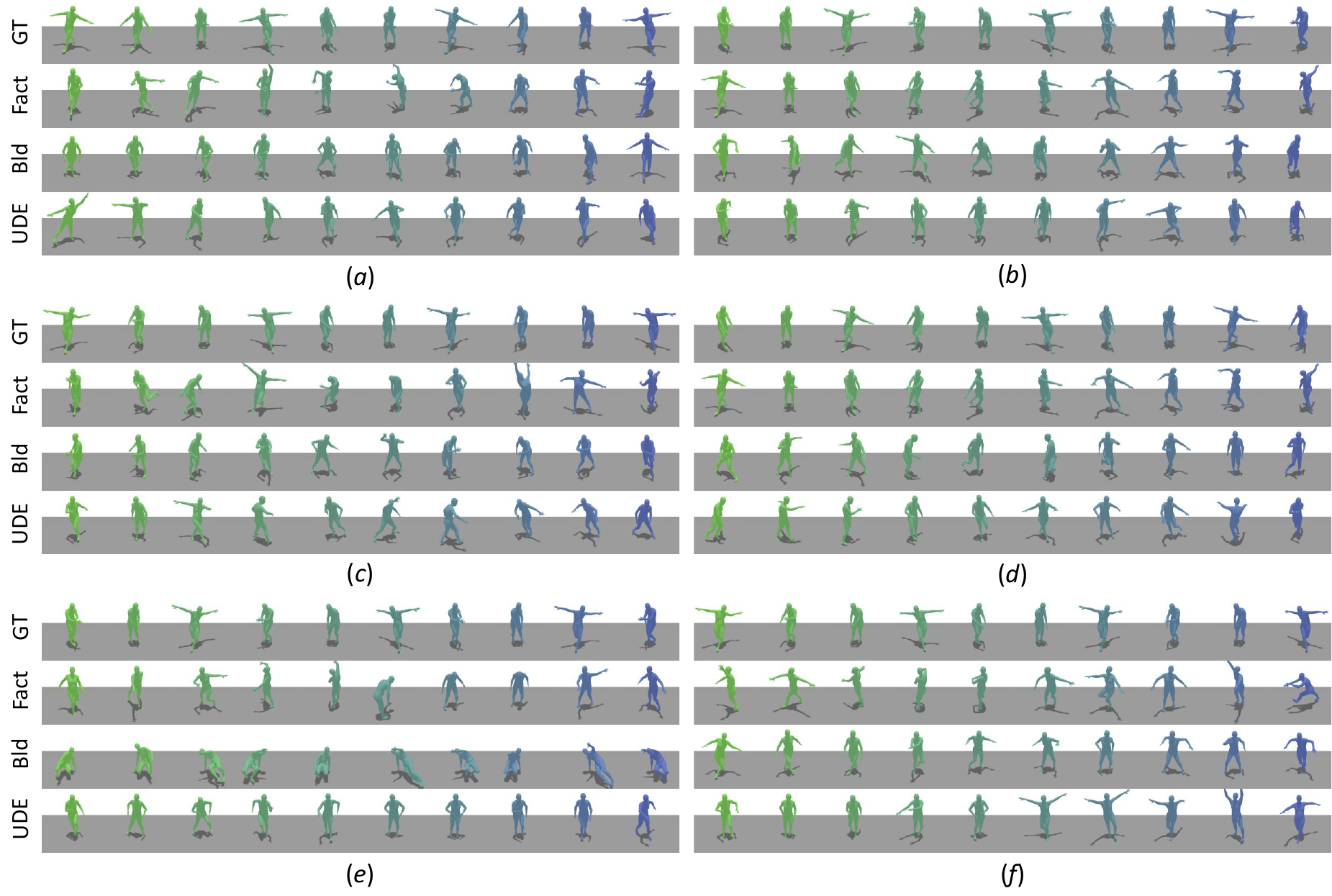}
    \caption{\textbf{Visual Comparison of Dance Moves Generated by \textit{BR} Style Music.} The average MDSC: Bailando $>$ UDE $\approx$ FACT.}
    \label{fig:supp-br}
\end{figure*}

\begin{figure*}
    \centering
    \includegraphics[width=0.88\textwidth]{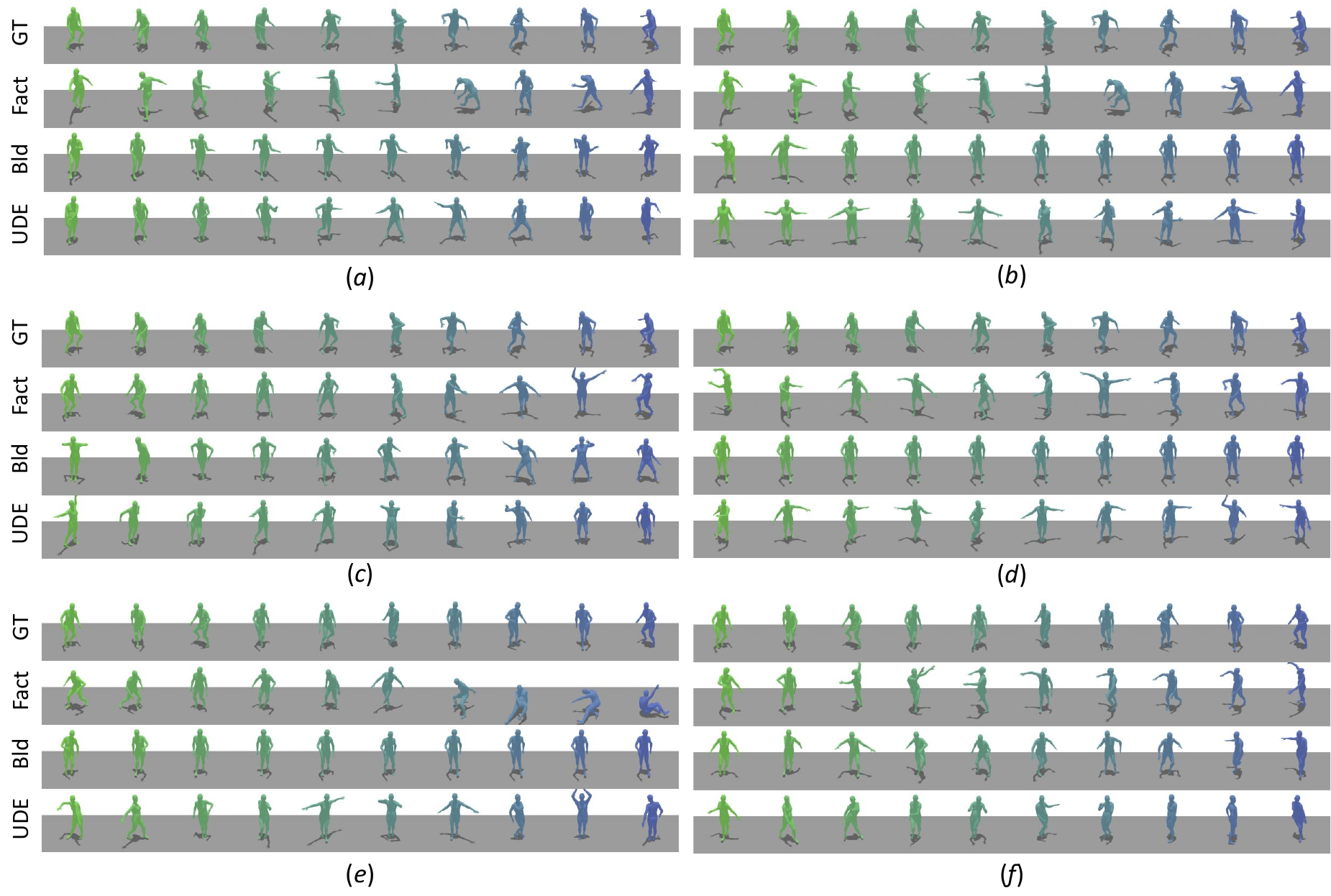}
    \caption{\textbf{Visual Comparison of Dance Moves Generated by \textit{HO} Style Music.} The average MDSC: Bailando $>$ FACT $>$ UDE.}
    \label{fig:supp-br}
\end{figure*}

\begin{figure*}
    \centering
    \includegraphics[width=0.88\textwidth]{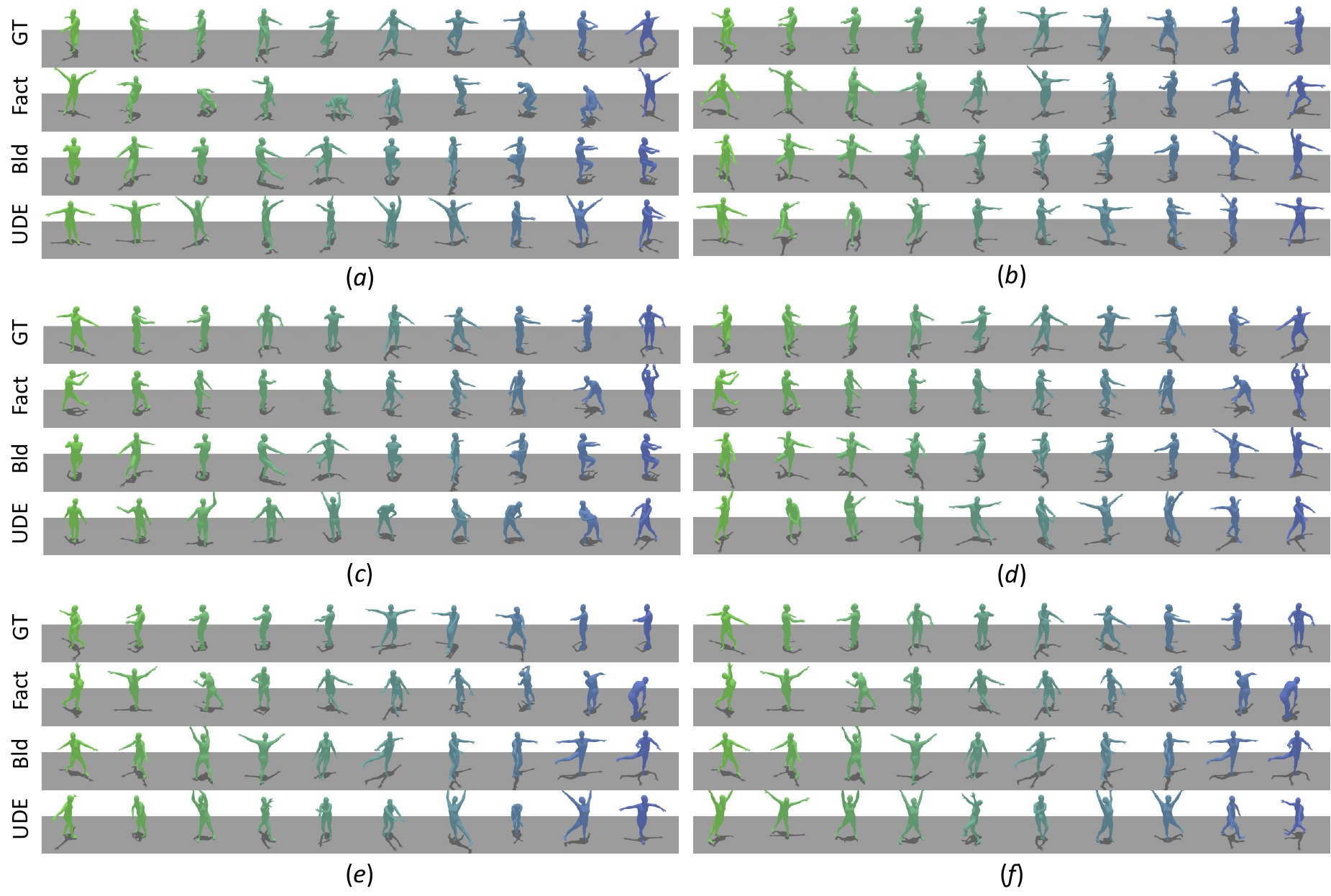}
    \caption{\textbf{Visual Comparison of Dance Moves Generated by \textit{JB} Style Music.} The average MDSC: Bailando $>$ FACT $>$ UDE.}
    \label{fig:supp-br}
\end{figure*}

\begin{figure*}
    \centering
    \includegraphics[width=0.88\textwidth]{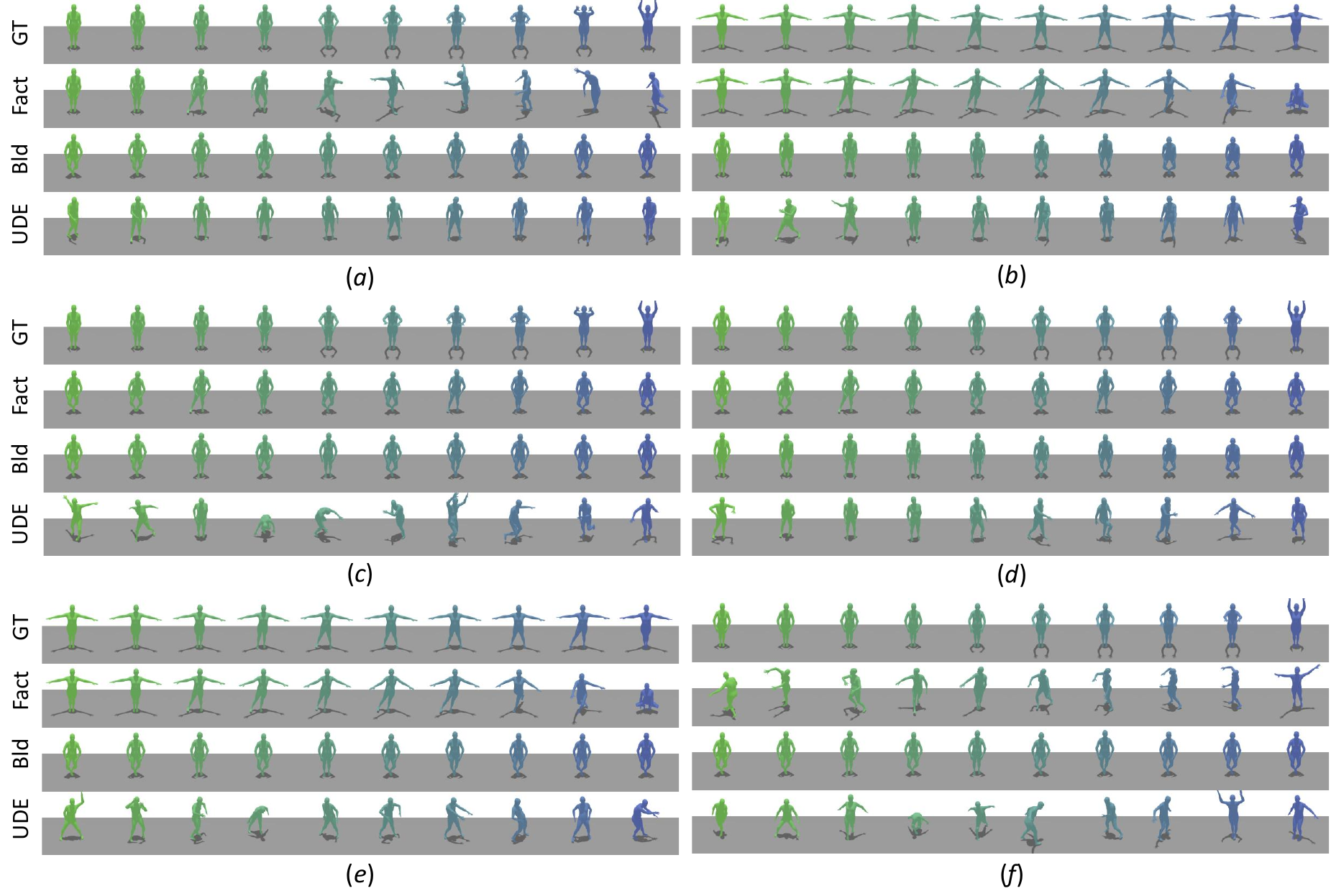}
    \caption{\textbf{Visual Comparison of Dance Moves Generated by \textit{JS} Style Music.} The average MDSC: Bailando $>$ FACT $>$ UDE.}
    \label{fig:supp-br}
\end{figure*}

\begin{figure*}
    \centering
    \includegraphics[width=0.88\textwidth]{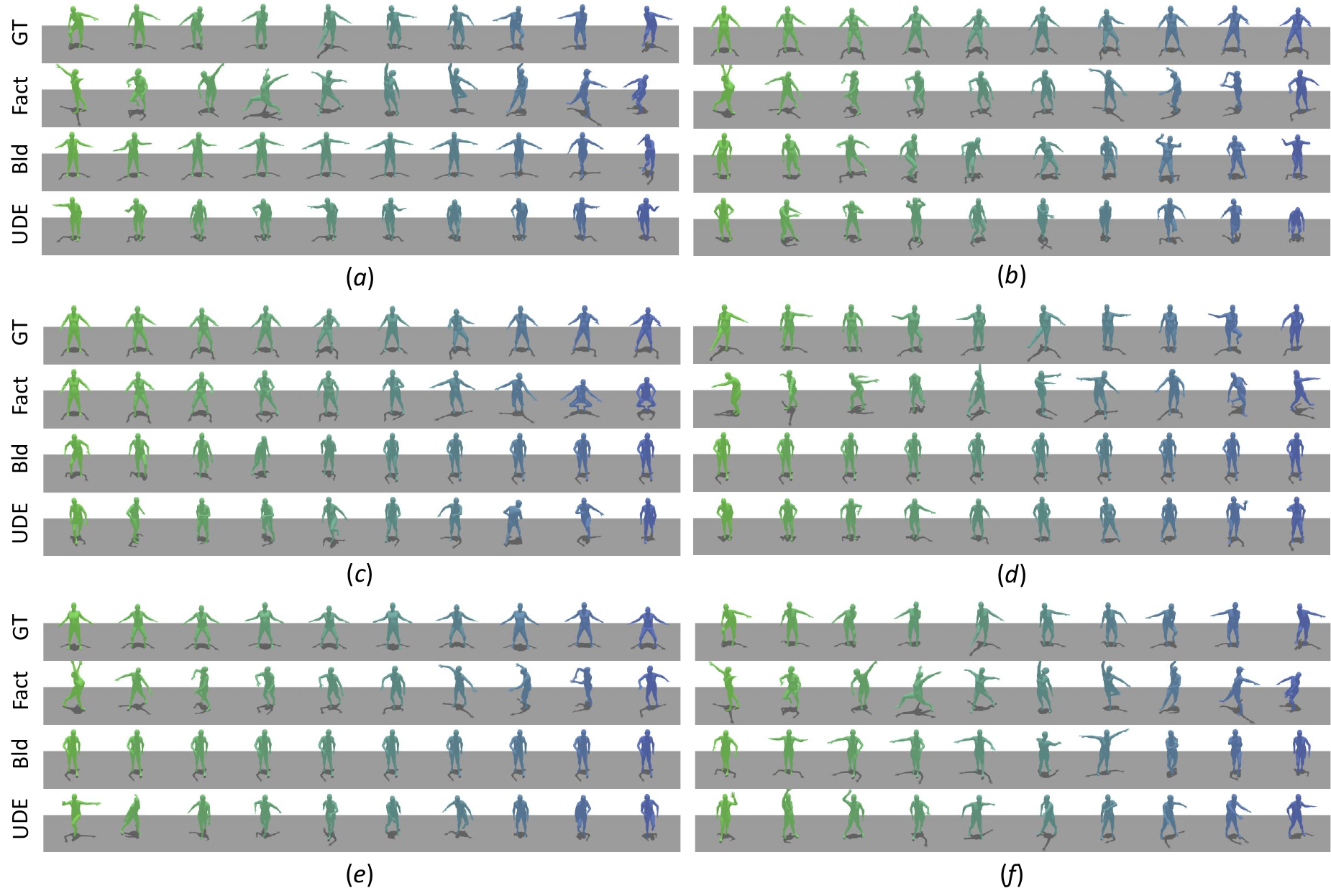}
    \caption{\textbf{Visual Comparison of Dance Moves Generated by \textit{KR} Style Music.} The average MDSC: Bailando $>$ FACT $>$ UDE.}
    \label{fig:supp-br}
\end{figure*}

\begin{figure*}
    \centering
    \includegraphics[width=0.88\textwidth]{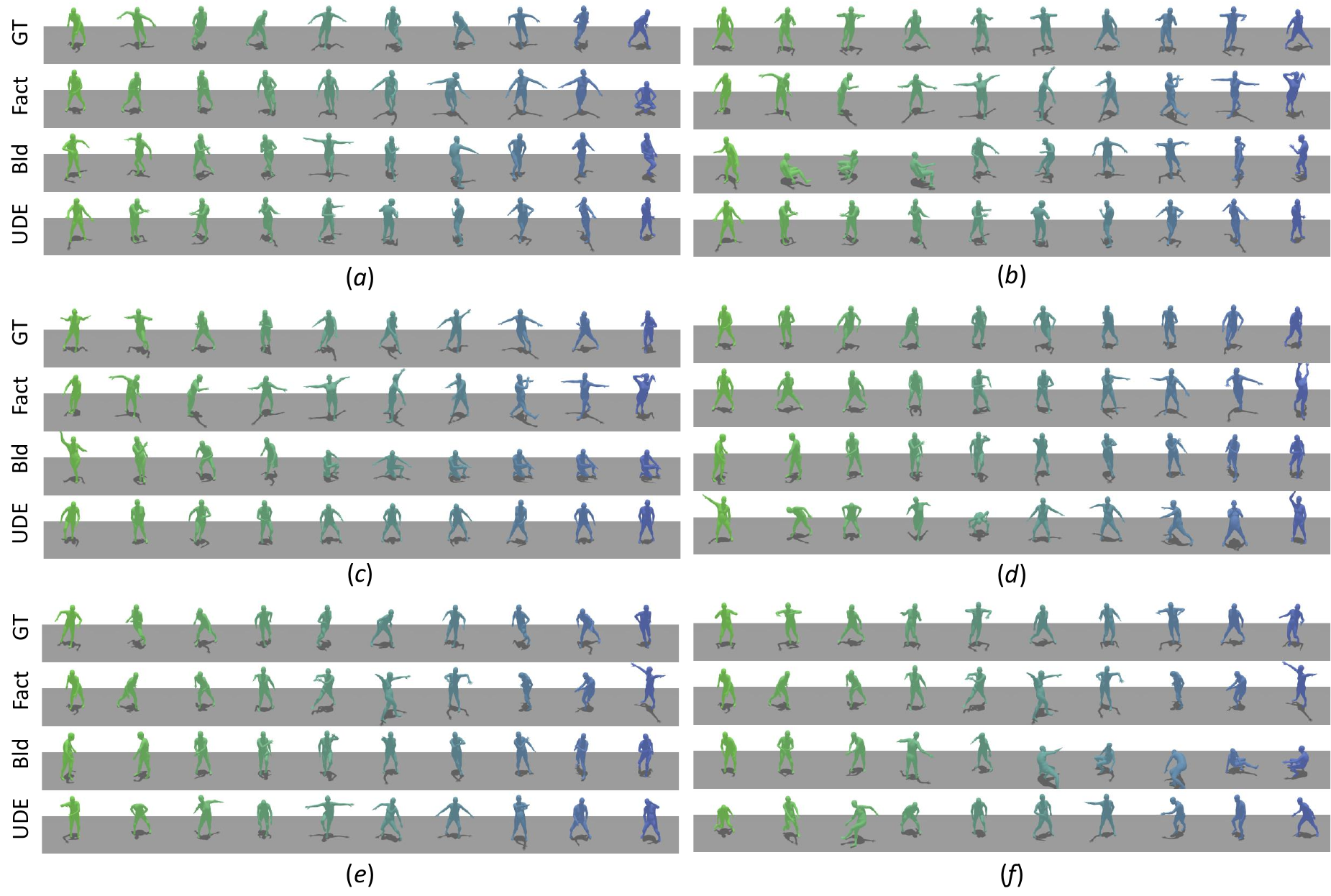}
    \caption{\textbf{Visual Comparison of Dance Moves Generated by \textit{LH} Style Music.} The average MDSC: UDE $>$ FACT $\approx$ Bailando.}
    \label{fig:supp-br}
\end{figure*}

\begin{figure*}
    \centering
    \includegraphics[width=0.88\textwidth]{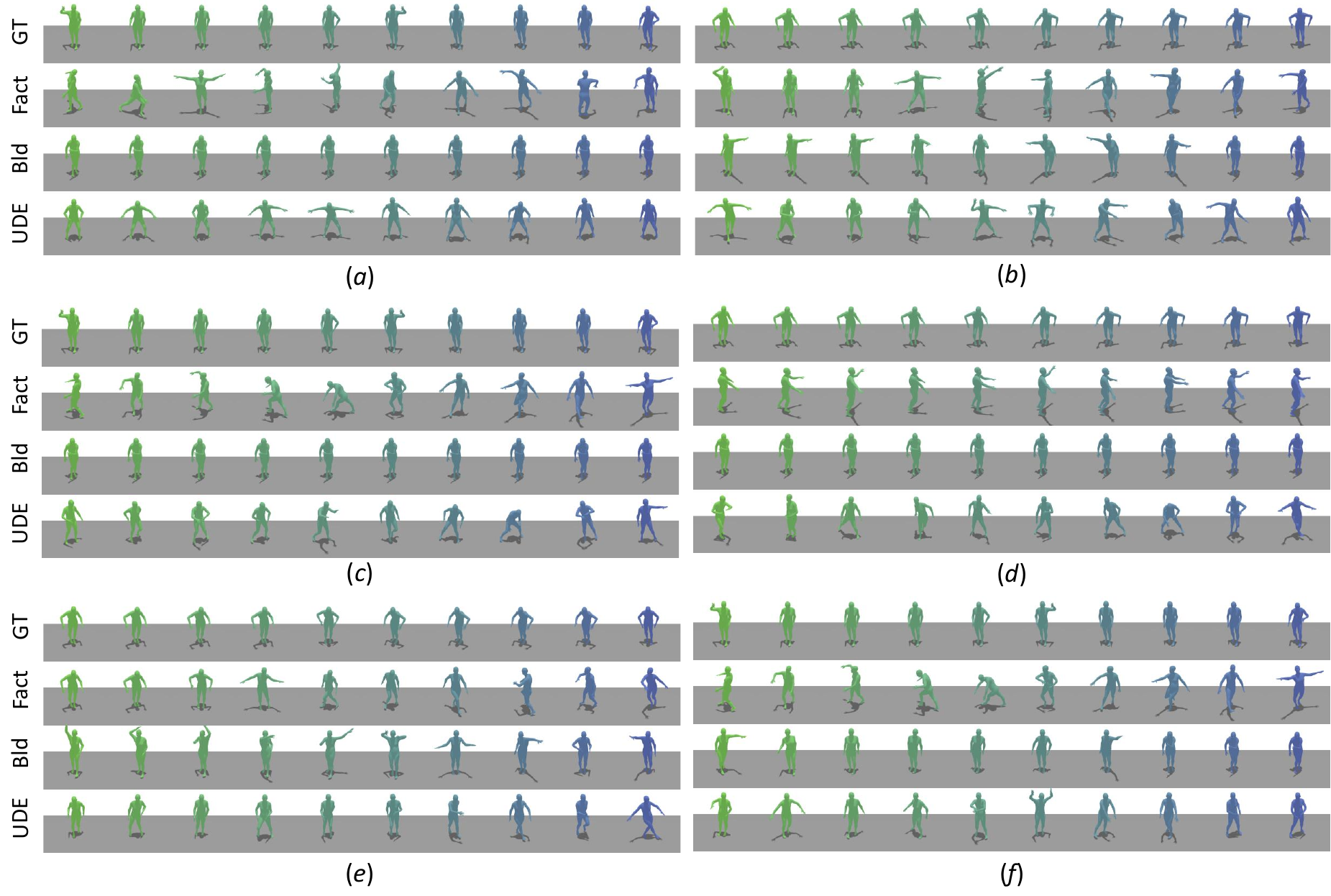}
    \caption{\textbf{Visual Comparison of Dance Moves Generated by \textit{LO} Style Music.} The average MDSC: Bailando $>$ FACT $>$ UDE.}
    \label{fig:supp-br}
\end{figure*}

\begin{figure*}
    \centering
    \includegraphics[width=0.88\textwidth]{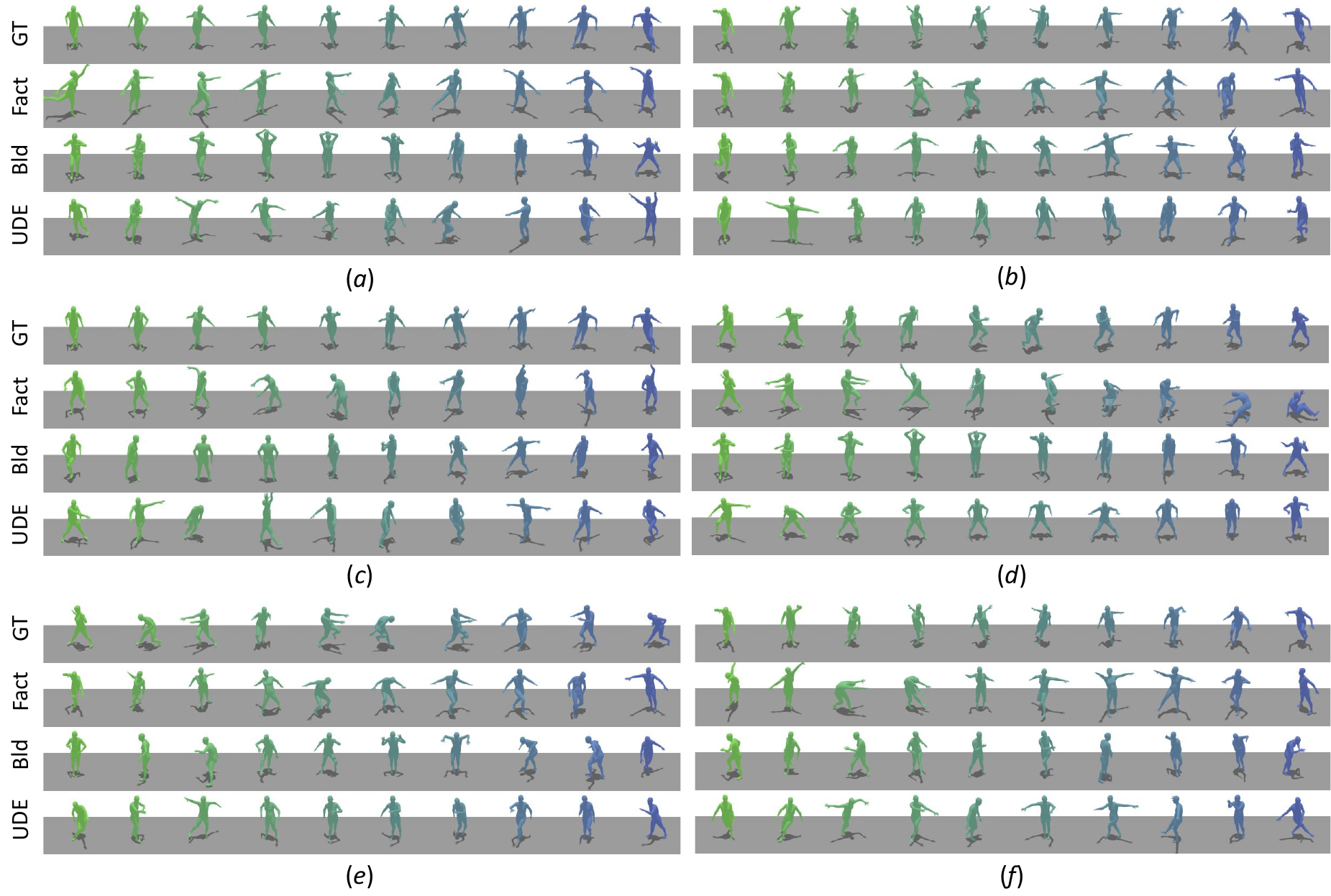}
    \caption{\textbf{Visual Comparison of Dance Moves Generated by \textit{MH} Style Music.} The average MDSC: UDE $>$ Bailando $>$ FACT.}
    \label{fig:supp-br}
\end{figure*}

\begin{figure*}
    \centering
    \includegraphics[width=0.88\textwidth]{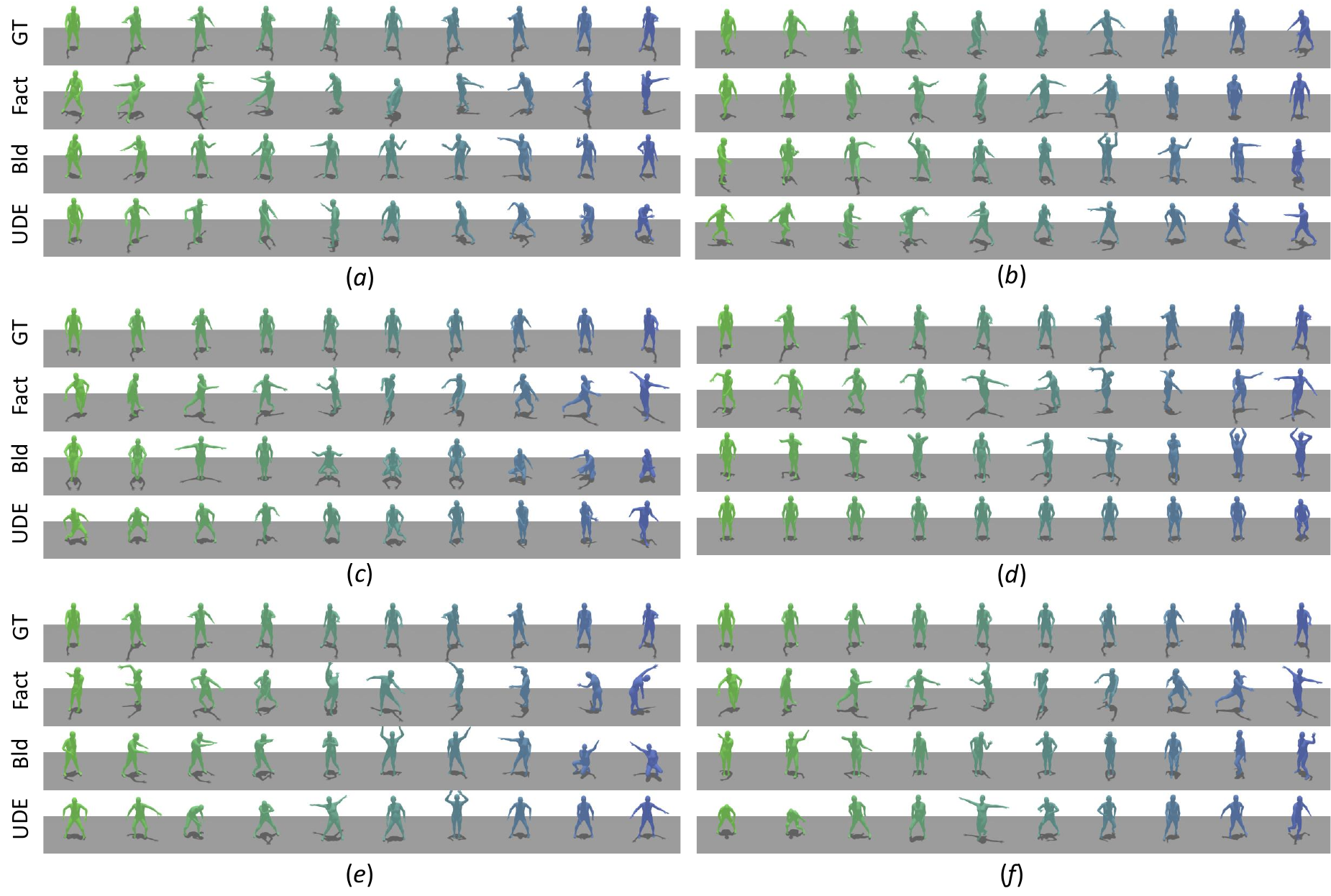}
    \caption{\textbf{Visual Comparison of Dance Moves Generated by \textit{PO} Style Music.} The average MDSC: UDE $\approx$ Bailando $>$ FACT.}
    \label{fig:supp-br}
\end{figure*}

\begin{figure*}
    \centering
    \includegraphics[width=0.88\textwidth]{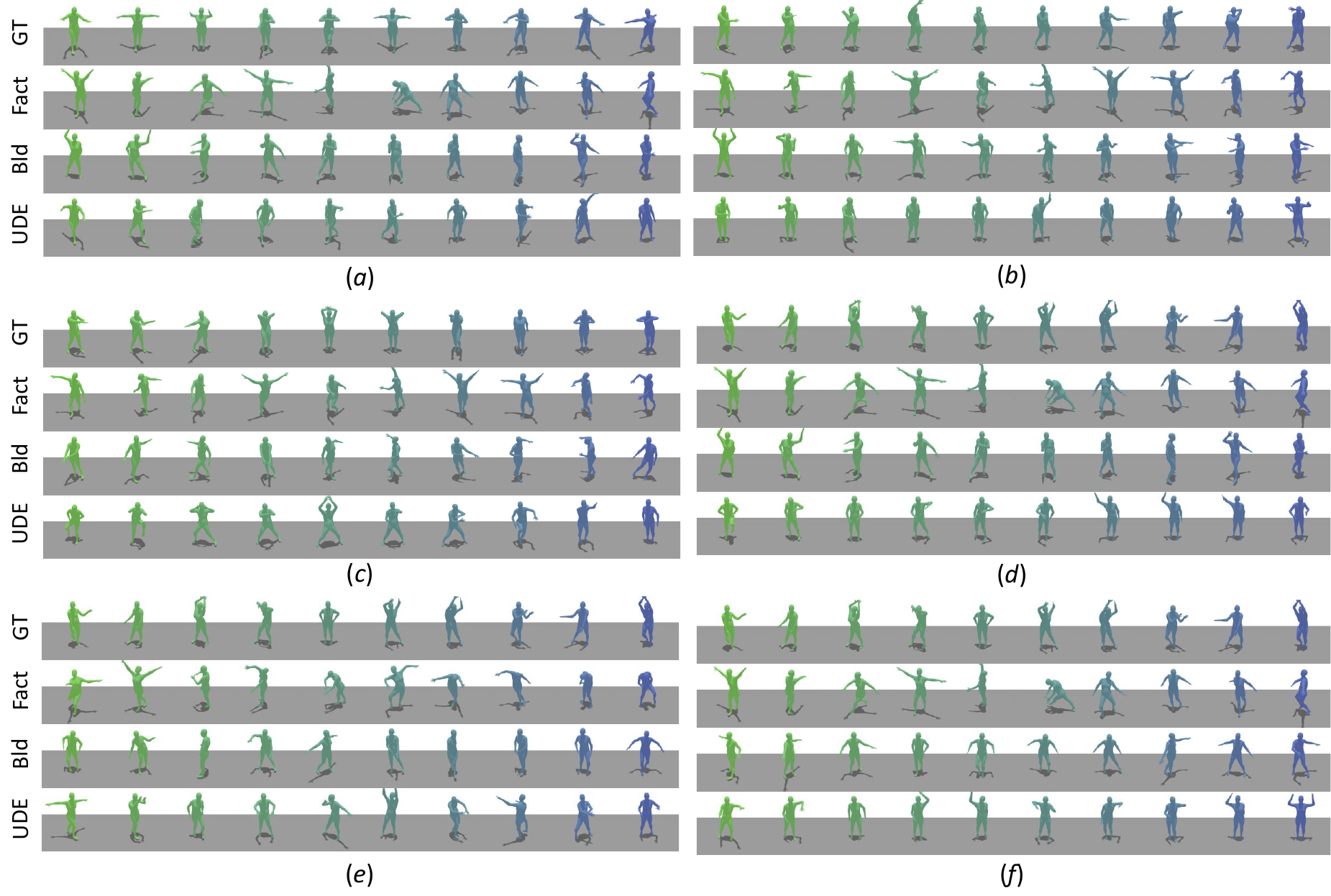}
    \caption{\textbf{Visual Comparison of Dance Moves Generated by \textit{WA} Style Music.} The average MDSC: FACT $\approx$ UDE $>$ Bailando.}
    \label{fig:supp-br}
\end{figure*}

% \clearpage
% {
%     \small
%     \bibliographystyle{ieeenat_fullname}
%     \bibliography{main}
% }

\end{document}